\documentclass[%
 reprint,
 amsmath,amssymb,
 aps,
]{revtex4-2}

\usepackage{graphicx}
\usepackage{dcolumn}
\usepackage{bm}
\usepackage{xcolor}

\begin{document}

\preprint{APS/123-QED}
\title{Supercurrent Distribution in Real-Space and Anomalous Paramagnetic Response in a Superconducting Quasicrystal}
\author{Takumi~Fukushima$^{1,2}$}
\email{tfukushima@issp.u-tokyo.ac.jp}
\author{Nayuta~Takemori$^{3,4}$}
\email{nayuta.takemori.qiqb@osaka-u.ac.jp}
\author{Shiro~Sakai$^4$}
\author{Masanori~Ichioka$^2$}
\author{Anuradha~Jagannathan$^{5}$}


\affiliation{
$^1$Institute for Solid State Physics, University of Tokyo, Kashiwa 277-8581, Japan\\
$^2$Research Institute for Interdisciplinary Science, Okayama University, Okayama 700-8530, Japan\\
$^3$Center for Quantum Information and Quantum Biology, Osaka University, Toyonaka, 560-0043, Japan\\
$^4$Center for Emergent Matter Science, RIKEN, Wako, Saitama 351-0198, Japan\\
$^5$Laboratoire de Physique des Solides, Universit\'{e} Paris-Saclay, 91405 Orsay, France
}

\date{\today}

\begin{abstract}

We theoretically study the real-space distribution of the supercurrent that flows under a uniform vector potential in a two-dimensional quasiperiodic structure. This is done by considering the attractive Hubbard model on the quasiperiodic Ammann-Beenker structure and studying the superconducting phase 
within the Bogoliubov-de Gennes mean-field theory. 
Decomposing the local supercurrent into the paramagnetic and diamagnetic components, we numerically investigate their dependencies on average electron density, temperature, and the angle of the applied vector potential.
We find that the diamagnetic 
current locally violates the current conservation law, necessitating compensation from the 
paramagnetic current, even at zero temperature. 
The paramagnetic current shows exotic behaviors in the quasiperiodic structure, such as local currents which are oriented transversally or 
reversely
to that of the applied vector potential.

\end{abstract}

\maketitle


\section{Introduction}
\label{sec:intro}

A quasicrystal is a solid that lacks translational symmetry but 
exhibits
a diffraction pattern with sharp Bragg peaks 
and 
a rotational symmetry
forbidden in periodic lattices~\cite{PhysRevLett.53.1951,qcs}. 
The quasicrystalline structure results in
exotic electronic states, such as critical 
states~\cite{Kohmoto1983,ostlund1983,Qian1986,tsunetsugu1986,critical,Tokihiro,tsunetsugu1}, which are distinct from those of conventional periodic crystals. 
Recently, an experimental work discovered bulk superconductivity in a Bergmann-type Al-Zn-Mg quasicrystalline alloy ($T_{\rm c}\sim50~{\rm mK}$) ~\cite{Kamiya2018}. More recently, superconductivity has also been reported in a van der Waals layered quasicrystal Ta-Te ($T_{\rm c} \sim 1~{\rm K}$)~\cite{tokumoto23}.
The superconductivity in quasicrystals poses new questions since they do not possess fundamental prerequisites such as the Fermi surface in the conventional Bardeen-Cooper-Schrieffer (BCS) theory~\cite{BCS} due to the absence of translational symmetry.
In previous theoretical works, such superconductivity has been studied by considering the attractive Hubbard model on quasiperiodic lattices. These studies showed that superconducting pairing is inhomogeneous, with real-space distributions of the site-dependent local electron density and superconducting order parameter~\cite{tezuka2010,tezuka2013,cai2013,fulga2016,Sakai2017,A-Bsuper,sakai2019,cao2020,Takemorisc,ghadimi2021,liu2022cooper,Uri23}. More interestingly, it has been pointed out by two of the present authors that non-BCS type superconductivity, comprised of Cooper pairs with finite center-of-mass momentum, exists in the weak-coupling region~\cite{Sakai2017}.

The interplay between the quasiperiodicity and superconductivity was studied in earlier works on quasiperiodic pinning arrays in periodic superconductors~\cite{Misko05,Misko06,Kemmler06,Silhanek06,Misko10}, as well as quasiperiodic networks of ordinary superconducting wires~\cite{Gordon86,Behrooz86,Springer87,Nori87,Nori88,Niu89}. 
These were studies for 
superconductivity in artificially fabricated quasiperiodic structures. 
In contrast, we investigate the electromagnetic response of superconducting quasicrystals at the atomic level.
In particular, supercurrent that flows
in response to a uniform vector potential such as the Meissner current 
is a basic property that has however been scarcely explored.
In a periodic system with a simple unit cell, it is 
obvious that the local supercurrents 
are uniformly distributed in the lattice, due to the homogeneity of the superconducting state in this case. 
In a simple crystal, each of the paramagnetic and diamagnetic components~\cite{schrieffer} of the local supercurrent 
is also uniform.
In contrast, we will show that in the quasicrystal, where both the local electron density and superconducting order parameter are spatially varying, 
the local supercurrent exhibits a nontrivial spatial dependence as well.  

In this study, we consider a two-dimensional quasiperiodic structure and the local supercurrent flow induced by an external uniform vector potential.
We investigate the attractive Hubbard model on the Ammann-Beenker structure~\cite{beenker,AB,ammann} by means of the Bogoliubov-de Gennes (BdG) mean-field theory.
First, we formulate the expression of the local supercurrent under the uniform vector potential and discuss its real-space distribution on the structure.
To clarify how the inhomogeneity of the superconducting state affects the supercurrent flow, we further decompose it into the paramagnetic and diamagnetic 
components.
We then discuss the dependence of the local supercurrent on
(i) average filling, (ii) temperature, and (iii) angle 
of the applied vector potential.
Interestingly, we find that the non-uniform diamagnetic current can locally violate the current conservation law, i.e. 
have a non-zero divergence. The paramagnetic current flows so as to re-establish the conservation of the local current. This leads to non-uniform supercurrent distributions which are unique to quasiperiodic systems. We find furthermore that the paramagnetic currents do not vanish at zero temperature, an anomalous property that was observed earlier by Liu {\it et al}.~\cite{liu2022cooper} in the site averaged value.

The rest of this paper is organized as follows. In Sec.~\ref{sec:method}, we introduce the model Hamiltonian and explain our theoretical approach. We discuss the distribution of the local supercurrent and its dependencies on the average filling, temperature, and the angle of the applied vector potential
in Sec.~\ref{sec:results}.
A brief summary is given in Sec.~\ref{sec:sum}.
The relation between formulations in our previous study~\cite{TF} and the present one is explained in 
Appendix.

\section{Model and Method}
\label{sec:method}

\begin{figure}[tb]
\includegraphics[width=0.4\textwidth]{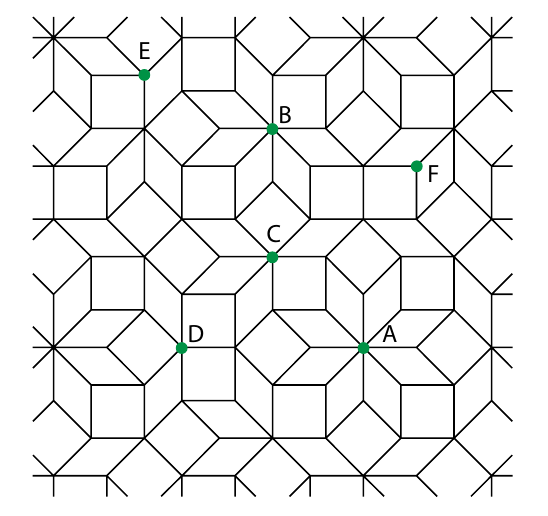}
\caption{\label{fig:ABtiling}A part of the Ammann-Beenker structure. The structure has six different vertex patterns, with A through F assigned in descending order of coordination number $Z_{i}=8, 7,\cdots,3$~\cite{Jag2005,Jag2004}.
}
\end{figure}

This study is carried out on the Ammann-Beenker structure (Fig.~\ref{fig:ABtiling}), which is a two-dimensional quasiperiodic tiling with an eight-fold rotational symmetry~\cite{beenker,AB,ammann}. For our numerical calculations, we 
use a square approximant of the perfect infinite tiling, consisting of $N=1393$ sites. 
This square approximant of the Ammann-Beenker structure was generated by the cut-and-project method~\cite{duneau}.
Here, we adopt a vertex model, where an atomic orbital is placed on each vertex of the Ammann-Beenker tiling.
The coordination number $Z_i$ at each site ranges from 3 to 8 and the vertices can be categorized into six classes if one does not distinguish two geometries of $Z_i=5$~\cite{Jag2005,Jag2004}. 

We consider an attractive ($U<0$) Hubbard model~\cite{Esslinger2010} to study $s$-wave superconductivity in this Ammann-Beenker structure, as was done in previous studies~\cite{Sakai2017, Takemorisc,A-Bsuper}.
To study the local supercurrent in the presence of a vector potential, we include ${\bm A}(\bm r)$ as the Peierls phase in the transfer term of the model Hamiltonian~\cite{peierls}. 
Thus, the Hamiltonian is given by
\begin{eqnarray}
\label{eq:hamiltonian} && 
      {\hat {\cal H}}=-\sum_{\langle i,j\rangle\sigma} \left\{
      t \exp \left(-{\mathrm i}\int_{\bm r_{j}}^{\bm r_{i}} {\bm A}(\bm r) \cdot {\mathrm d}{\bm r} \right) {\hat c}^{\dag}_{i\sigma}{\hat c}_{j\sigma}+h.c. \right\} 
      \nonumber \\  && \hspace{2cm}
+U\sum_{i}\hat{n}_{i\uparrow}\hat{n}_{i\downarrow}-\mu\sum_{i\sigma}\hat{n}_{i\sigma}.
\end{eqnarray}
Here, ${\hat c}^{\dag}_{i\sigma}$ (${\hat c}_{i \sigma}$) creates (annihilates) an electron of spin $\sigma$ at site $i$. 
We suppose a finite electron-transfer integral $t$ only between the nearest neighbor sites (denoted by $\langle i,j \rangle$) connected by an edge of a square or a rhombus and set it as the unit of energy.
In the noninteracting limit, the energy width of the 
density of states is about $8.5t$~\cite{sakai2021}. 
We define a local electron density $n_i=\sum_\sigma \langle \hat {n}_{i\sigma} \rangle$ with $ {\hat n}_{i\sigma}={\hat c}_{i\sigma}^{\dagger}{\hat c}_{i \sigma}$.
The chemical potential $\mu$ is tuned to fix the average electron density ${\bar n}=\sum_{i}n_{i}/N$ where $N$ is the system size. 
We fix the attractive interaction strength to $U=-3$, and select the averaged electron density ${\bar n}=0.3$, 0.5, 0.7, and 0.9 to avoid the delta-function singularity in the density of states at the half-filling due to confined states \cite{KogaAB}.

In the case of a uniform vector potential ${\bm A}$, the Peierls phase in Eq. (\ref{eq:hamiltonian}) can be rewritten as 
\begin{eqnarray}
 -{\rm i}\int_{{\bm r}_{j}}^{{\bm r}_{i}}{\bm A}(\bm r) \cdot {\rm d}{\bm r}
  =-{\rm i}{\bm A} \cdot {\bm r}_{ij} , 
\end{eqnarray}
where
${\bm r}_{ij}={\bm r}_{i}-{\bm r}_{j}
= a {\bm e}_n$ is
the bond vector between sites
$i$ and $j$. Here $a$ is the bond length and the unit vectors ${\bm e}_n = (\cos \phi_n, \sin \phi_n)$ with $\phi_n=2 n \pi/8$ ($n=0, \pm 1, \pm 2, \pm 3, 4$) correspond to the eight permitted bond orientations on the tiling, where we take $\phi_n=0$ as the $x$ direction.

We henceforth assume that the uniform vector potential
${\bm A}$ 
is applied parallel to the plane of the Ammann-Beenker structure and evaluate the local supercurrent induced by 
$\bm A$ on each bond ${\bm r}_{ij}$. 
We control the direction of the vector potential 
${\bm A}= -|{\bm A}|(\mathrm{cos}\theta,\mathrm{sin}\theta)$ by changing the angle parameter $\theta$  
in the range $0\leq \theta < \frac{\pi}{4}$. 

Using the mean-field approximation, Hamiltonian in Eq. (\ref{eq:hamiltonian}) is reduced to 
\begin{eqnarray} 
\label{eq:Hmf-1}
{\hat{\cal H}} =
\sum_{i,j}
\left(\begin{array}{cc}
   {\hat c}^\dagger_{i\uparrow} & 
   {\hat c}_{i\downarrow}\end{array}\right)
\hat{\cal H}_{i,j}
\left(\begin{array}{c}
   {\hat c}_{j\uparrow}\\
   {\hat c}^{\dag}_{j\downarrow}\end{array}\right)
\end{eqnarray}
with 
\begin{eqnarray} 
\label{eq:Hmf-2}
\hat{\cal H}_{i,j}
=\left( \begin{array}{cc} 
     K_{\uparrow i,j }  & \Delta_i \delta_{i,j} \\
     \Delta_i^\ast \delta_{i,j} &  - K^\ast_{ \downarrow i,j} 
\end{array} \right)
\end{eqnarray}
where $K_{\sigma i,j}=-t\delta_{\langle i,j \rangle}{\mathrm{e}}^{-{\rm i}{\bm A} \cdot {\bm r}_{ij}}+(U n_{i\bar\sigma}-\mu)\delta_{i,j}$ is a kinetic term with $n_{i \sigma}=\langle \hat{n}_{i \sigma} \rangle$ and $\Delta_{i}=-U\langle {\hat c}_{i\uparrow}{\hat c}_{i\downarrow} \rangle$ is the site-dependent superconducting order parameter. 
$\delta_{\langle i,j \rangle}$ is the Kronecker delta that counts only between the nearest neighbor sites~\cite{Takemorisc}. 
We note that 
the Hartree term $Un_{i {\bar \sigma}}$ needs to be explicitly incorporated above because it has a site dependence and hence cannot be absorbed into the chemical potential term~\cite{Sakai2017, Takemorisc,A-Bsuper,dobrosavljevic,kita}.
The Hamiltonian of $2N \times 2N$ matrix in Eq.~(\ref{eq:Hmf-1}) 
is diagonalized through the Bogoliubov transformation 
\begin{eqnarray}
\left(\begin{array}{c}
   {\hat c}_{i\uparrow}\\
   {\hat c}^{\dag}_{i\downarrow} \\
   \end{array} \right)
   =\sum_{\epsilon}
   \left(\begin{array}{cc}
       u_{\epsilon}({\bm r}_{i}) & -v^{*}_{\epsilon}({\bm r}_{i}) \\
        v_{\epsilon}({\bm r}_{i})&u^{*}_{\epsilon}({\bm r}_{i})  \\
   \end{array}\right)
   \left( \begin{array}{c} \notag
     {\hat \gamma}_{\epsilon \uparrow} \\ 
     {\hat \gamma}^{\dag}_{\epsilon \downarrow}\\
   \end{array} \right)     
\end{eqnarray}
and we obtain the Bogoliubov-de Gennes (BdG) equation~\cite{de2018superconductivity,Nagai2020,Takemorisc,PhysRevB.106.064506,TF} 
\begin{eqnarray}
\label{eq:BdG}
  \sum_{j}
\hat{\cal H}_{i,j} 
   \left( \begin{array}{c}
     u_{\epsilon}({\bm r}_j) \\ 
     v_{\epsilon}({\bm r}_j)\\
   \end{array} \right) 
 = E_\epsilon 
   \left( \begin{array}{c}
     u_{\epsilon}({\bm r}_i) \\ 
     v_{\epsilon}({\bm r}_i) \\ 
   \end{array} \right).
\end{eqnarray}
Here, $E_{\epsilon}$ denotes an eigenenergy of the BdG Hamiltonian $\hat{\cal H}_{i,j}$ 
and $u_{\epsilon}(\bm{r}_{i}), v_{\epsilon}(\bm{r}_{i})$ denote wave functions on the site $i$. The index $\epsilon$ distinguishes eigenstates of the BdG  Hamiltonian, which runs over 1 to $2N$, including eigenstates with both positive and negative $E_{\epsilon}$. 
As the self-consistent condition, 
the gap equation and local electron density for each spin are obtained as~\cite{PhysRevB.96.094522,PhysRevB.65.014508}
\begin{eqnarray}
\label{eq:gap-eq}
  \Delta_{i}=-U\sum_{\epsilon} u_{\epsilon}({\bm r}_{i})v^{\ast}_{\epsilon}({\bm r}_{i})(1-f(E_{\epsilon})),
\end{eqnarray}
\begin{eqnarray}
\label{eq:density1-eq}
  n_{i \uparrow}=\sum_\epsilon |u_{\epsilon}
  ({\bm r}_{i})|^2f(E_{\epsilon}),
\end{eqnarray}
\begin{eqnarray}
\label{eq:density2-eq}
  n_{i \downarrow}=\sum_\epsilon |v_{\epsilon}({\bm r}_{i})|^2(1-f(E_{\epsilon})),
\end{eqnarray}
with $f(E)=1/(e^{E/T}+1)$ is the Fermi-Dirac distribution function at temperature $T$.
Using only positive $E_{\epsilon}$, we obtain the conventional formula 
$\Delta_i=-U\sum^{N}_{\epsilon=1} u_{\epsilon}({\bm r}_{i})v^{\ast}_{\epsilon}({\mathbf r}_{i})(1-2f(E_{\epsilon}))$
\cite{de2018superconductivity}, where we have used the particle-hole symmetry, 
$(u_{\epsilon}({\mathbf r}_i),v_{\epsilon}({\mathbf r}_i))\rightarrow
(-v^{*}_{\epsilon}({\mathbf r}_i),u^{*}_{\epsilon}({\mathbf r}_i))$ 
as 
$E_{\epsilon}\rightarrow-E_{\epsilon}$, for a negative $E_{\epsilon}~(\epsilon\in[N+1,2N])$ in Eq.~(\ref{eq:gap-eq}). 

The local supercurrent from a site $j$ to $i$ is given by 
\begin{eqnarray}
\label{eq:Jtot1} && 
{\bm J}_{j \rightarrow i} 
=-\frac{\partial{\langle{\hat{\cal  H}(\bm A)}\rangle}}{\partial {\bm A}}
\\ && 
\label{eq:Jtot2}
=2t {\rm Im}\left\{ 
     \exp(-{\rm i}{\bm A}\cdot {\bm r}_{ij})\sum_{\sigma}\langle{\hat{c}^{\dag}_{i\sigma}\hat{c}_{j\sigma}\rangle}
     \right\}  {\bm r}_{ij}. 
\nonumber \\ &&
\end{eqnarray}
It can be divided into the paramagnetic current 
\begin{eqnarray}
\label{eq:Jpara}
{\bm J}^{\mathrm{para}}_{j \rightarrow i} 
=2t \cos \left({\bm A}\cdot {\bm r}_{ij} \right)\mathrm {Im}\left\{ \sum_{\sigma}\langle{\hat{c}^{\dag}_{i\sigma}\hat{c}_{j\sigma}\rangle} \right\}
{\bm r}_{ij},
\quad 
\nonumber \\ && 
\end{eqnarray}
and the diamagnetic current
\begin{eqnarray}
\label{eq:Jdia}
{\bm J}^{\mathrm{dia}}_{j \rightarrow i}
=-2t \sin \left({\bm A}\cdot {\bm r}_{ij} \right)
\mathrm {Re}\left\{ \sum_{\sigma}\langle{\hat{c}^{\dag}_{i\sigma}\hat{c}_{j\sigma}\rangle} \right\} 
        {\bm r}_{ij}
        \quad 
        \nonumber \\ && 
\end{eqnarray}
so that 
${\bm J}_{j \rightarrow i}
={\bm J}^{\mathrm{para}}_{j \rightarrow i}+{\bm J}^{\mathrm{dia}}_{j \rightarrow i}$.
Here, we have defined
the paramagnetic (diamagnetic) component as an even (odd) function of ${\bm A}$ in the expression of the local supercurrent. 
In the weak limit of the vector potential, trigonometric functions in Eqs.~(\ref{eq:Jpara}) and (\ref{eq:Jdia}) can be reduced to 1 and 
${\bm A}\cdot{\bm r}_{ij}=|{\bm A}||{\bm r}_{ij}|\cos\alpha$, respectively reproducing the conventional definition of 
${\bm J}^{\mathrm{dia}}_{j \rightarrow i}$ and ${\bm J}^{\mathrm{para}}_{j \rightarrow i}$ used in previous studies~\cite{liu2022cooper}. 
The angle parameter $\phi_n$ of ${\bm r}_{ij}$ specifies the flow direction of the local current.
Here, we define an angle $\alpha = \theta - \phi_n$ between the applied vector potential and the bond vector and call $\cos\alpha$ a bond factor.
We note that
${\rm Re} \{ \langle{\hat{c}^{\dag}_{i\sigma}\hat{c}_{j\sigma}\rangle} \}
= (\langle\hat{c}^{\dag}_{i\sigma}\hat{c}_{j\sigma}\rangle 
+\langle\hat{c}^{\dag}_{j\sigma}\hat{c}_{i\sigma}\rangle )/2$ represents the effective bond strength between the site $i$ and $j$ $(\neq i)$, and
${\rm Im} \{ \langle{\hat{c}^{\dag}_{i\sigma}\hat{c}_{j\sigma}\rangle} \}
= (\langle\hat{c}^{\dag}_{i\sigma}\hat{c}_{j\sigma}\rangle
-\langle\hat{c}^{\dag}_{j\sigma}\hat{c}_{i\sigma}\rangle)/2{\rm i}$
gives the net transfer from the site $j$ to $i$. 
In Eqs.~(\ref{eq:Jtot2})$-$(\ref{eq:Jdia}),
$\langle \hat{c}^{\dag}_{i \sigma}\hat{c}_{j \sigma} \rangle$ is obtained from the eigenstate of the BdG equation~(\ref{eq:BdG}) as 
\begin{eqnarray} 
    \langle \hat{c}^{\dag}_{i \uparrow}\hat{c}_{j \uparrow} \rangle = 
    \sum_{\epsilon}u^{*}_{\epsilon}({\bm r_{i}})u_{\epsilon}({\bm r_{j}})f(E_{\epsilon}),
\end{eqnarray}    
\begin{eqnarray}   
    \langle \hat{c}^{\dag}_{i \downarrow}\hat{c}_{j \downarrow} \rangle = 
    \sum_{\epsilon}v_{\epsilon}({\bm r_{i}})v^{*}_{\epsilon}({\bm r_{j}})(1-f(E_{\epsilon})).
\end{eqnarray}

In this study, we have chosen the amplitude of the uniform vector potential to have the value $|{\bm A}|=0.005$.
Since the local supercurrents are linear as a function of the vector potential in the weak $|{\bm A}|$ limit, changing the value of the external vector potential will not result in qualitative changes in our results.

\section{Results}
\label{sec:results}

\subsection{Real-space distribution of the local electron density and the superconducting order parameter}

\begin{figure*}[tb]

\centering
\includegraphics[width=\textwidth]{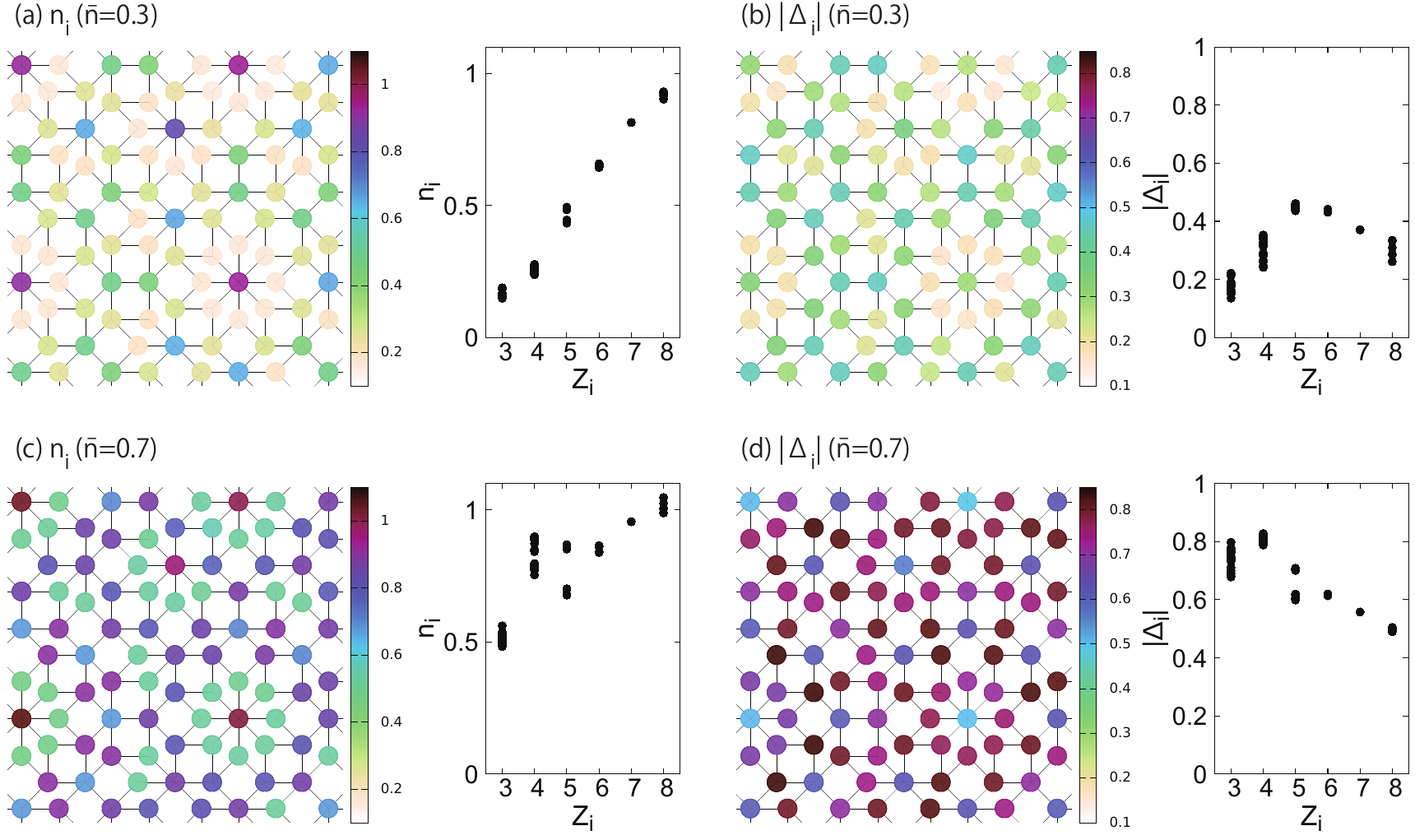}






   \caption{\label{fig:n-op}
  Real-space distribution (left panels) and the coordination number $Z_i$ dependence (right panels) of 
  the local electron density $n_i$ [(a) and (c)] and the superconducting order parameter amplitude $|\Delta_i|$ [(b) and (d)] on the Ammann-Beenker structure for ${\bar n}=0.3$ [(a) and (b)] and 0.7
  [(c) and (d)] for $U=-3$, $T=0.01$, and $\theta=0$. For the spatial distribution, 
  we show a part of the system consisting of about 100 sites for visibility.
}
 \end{figure*}

We begin by considering  the case when the uniform vector potential ${\bm A}$ is parallel to the $x$-axis, i.e., ${\bm  A} =(-|{\bm A}|,0)$ by setting $\theta=0$.
Before discussing the supercurrent distribution, we 
study the inhomogeneous distribution of local quantities.
Figure~\ref{fig:n-op} presents real-space distribution of the local electron density $n_i$ and the superconducting order parameter amplitude $|\Delta_i|$ after the self-consistent calculation of Eqs.~(\ref{eq:BdG})-(\ref{eq:density2-eq}) at $T=0.01$ for the fillings ${\bar n}=0.3$ and 0.7. 
The figure 
zooms in a small region consisting of about 100 sites, for which one clearly sees inhomogeneous spatial distributions of $n_i$ and $|\Delta_i|$, which moreover exhibit an approximate
eight-fold symmetry as reported in Ref.~\cite{A-Bsuper}. 
Classifying the vertices by the coordination number $Z_i$,
we plot the distributions of $n_i$ and $|\Delta_i|$ against $Z_i$ in the right panels. 
We note that the values of $n_i$ and $|\Delta_i|$ have variations even among the sites with the same coordination number since such sites have different next-nearest neighbor (or further neighbor) configurations.

For the filling ${\bar n}=0.3$, the right-hand panel of Fig.~\ref{fig:n-op}(a) shows that 
$n_i$ tends to increase with 
$Z_i$. This behavior can be deduced from a property of the non-interacting model: when the Fermi energy lies below the main pseudogap, it is the sites of large $Z_i$ which are preferentially occupied \cite{aj1994}. This leads, in the BdG equation in Eq.~(\ref{eq:density1-eq}), to the factors $|u_{\epsilon}
({\bm r}_{i})|^2$ being 
larger for larger
$Z_i$ values.
In contrast, the local order parameter amplitude, shown in Fig.~\ref{fig:n-op}(b) does not increase monotonically with $Z_i$ but has
a maximum at $Z_i=5$, as can be seen from the right-hand panel. This maximum in $|\Delta_i|$ can also be simply explained -- according to Eq.~(\ref{eq:gap-eq}), the order parameter amplitude is given by the product of $|u_{\epsilon}
  ({\bm r}_{i})|$ and $|v_{\epsilon}
  ({\bm r}_{i})|$, which are increasing and decreasing functions of $Z_i$ respectively. This leads to the maximum at $Z_i=5$, which is the value separating low and high coordination sites in this tiling.
As the filling is increased to larger values, for ${\bar n}=0.7$, the $n_i$ 
increases at all the sites, however, the differential increase is largest 
at the sites with smaller $Z_i$, as shown in Fig.~\ref{fig:n-op}(c). 
Overall, the distribution range of $n_i$ for different $Z_i$ becomes narrower as filling $\bar{n}$ is increased, until at half-filling one reaches the uniform state  $n_i=1$ for all sites.
For ${\bar n}=0.7$, the local superconducting order parameter is enhanced compared to the case of filling ${\bar n}=0.3$. As shown in the right-hand panel of Fig.~\ref{fig:n-op}(d), the values of $\Delta_i$ are in this case largest on the sites with $Z_i=3$ and 4. This reflects the fact that, for a large 
filling, the non-interacting local density of states at these sites is significantly larger than that of the sites of large $Z_i$ \cite{aj1994}, for reasons discussed in \cite{aj2023}. 
These site-dependences of $n_i$ and $|\Delta_i|$, which hold even in the  absence of the uniform vector potential ${\bm A}$, affect behaviors of local supercurrent flow as we now discuss below.

\begin{figure*}[tb]

\centering
\includegraphics[width=\textwidth]{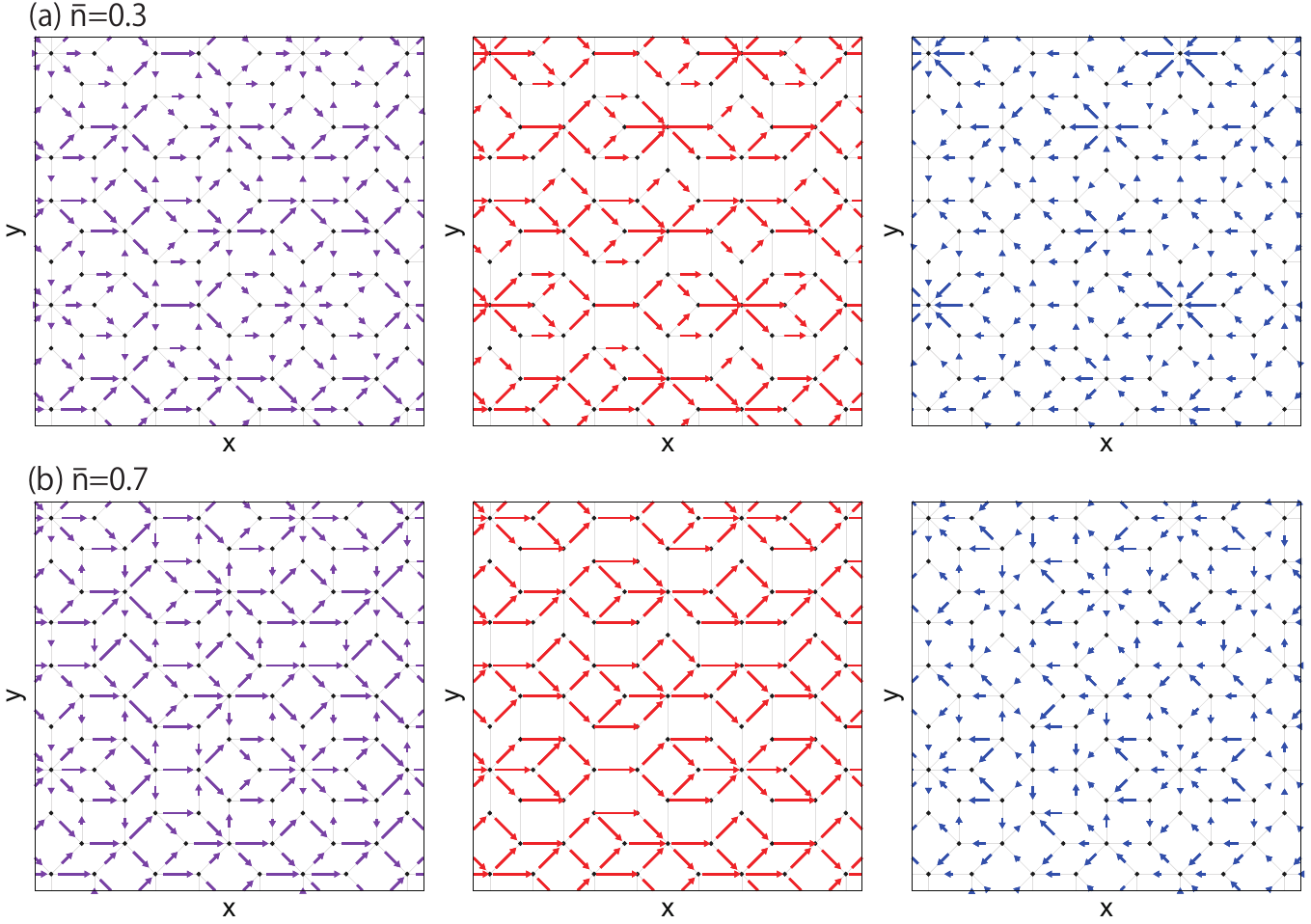}













\caption{
  \label{fig:xyJ}
  Real-space distribution of the local current ${\bm J}_{j \rightarrow i}$(left), diamagnetic current ${\bm J}^{\rm dia}_{j \rightarrow i}$ (middle) and paramagnetic current ${\bm J}^{\rm para}_{j \rightarrow i}$ (right) on the Ammann-Beenker structure at $U=-3$, $T=0.01$, and $\theta =0$. The results were obtained for $\bar n=0.3$ (a) and 0.7 (b). 
Length and orientation of arrows 
represent the
strength and direction of the supercurrent on each bond. 
Black dots show position of the vertex.
In each panel, we show a part of the system consisting of about 100 sites for visibility.
}
\end{figure*}



\subsection{Real-space distribution of the local supercurrent}
\label{subsec:realspaceJ}

Figure \ref{fig:xyJ} shows the spatial distribution of the supercurrent in the case of $\theta=0$.
Bonds are either parallel to this direction
($\phi_n=0$), perpendicular ($\phi_n=\frac{\pi}{2}$), or at an angle of $\frac{\pi}{4}$, which leads to large differences in the bond factor.
In uniform systems such as a square lattice, the differences in 
$J_{j \rightarrow i}=|{\bm J}_{j \rightarrow i}|$
among the bonds are attributed only to the bond factor $\cos\alpha$ 
for fixed $|{\bm A}|$.
It is just because 
${\bm J}_{j \rightarrow i} \sim {\bm J}^{\rm dia}_{j \rightarrow i}$
at sufficiently low temperature.
Therefore, the local supercurrent flows on the bonds where the bond factor $\cos\alpha$ is non-zero, i.e., the bonds with  $|\alpha|\neq\frac{\pi}{2}$.
Also, 
the same current flows for the same $\alpha$ bonds, forming a 
one-dimensional flow
distribution consisting of the respective local currents.
These are well-known responses of uniform superconductors~\cite{schrieffer}.

On the other hand, the local supercurrent ${\bm J}_{j \rightarrow i}$ in the inhomogeneous 
superconductor
flows non-uniformly as shown in the left panel of Fig.~\ref{fig:xyJ}(a) for ${\bar n}=0.3$, which is not determined simply by the bond factor. 
The overall tendency to 
flow along
one-dimensional ``channels" is similar to that of the uniform systems.
These one-dimensional channels 
having a cross-sectional width 
of a few lattice spacings are stacked along the $y$-direction. 
Notably, $J_{j \rightarrow i}$ 
depends on 
the sites
$i$ and $j$ even if the bonds have the same bond factor.
In addition, there are small supercurrent flows even in the directions of $|\phi_n|=\frac{\pi}{2}$. 
These features are characteristic of the quasiperiodic superconductor.

To 
understand the non-uniform distribution,
 we decompose ${\bm J}_{j \rightarrow i}$ into the diamagnetic current ${\bm J}^{\rm dia}_{j \rightarrow i}$
and paramagnetic current ${\bm J}^{\rm para}_{j \rightarrow i}$ as shown in the middle and right panels of Fig.~\ref{fig:xyJ}(a).
Since ${\bm J}^{\rm dia}_{j \rightarrow i}$ can be considered as a direct response to the vector potential ${\bm A}$ and has the bond factor $\cos\alpha$,
$J^{\rm dia}_{j \rightarrow i}$
of $\phi_n=0$ is larger than that of $|\phi_n|=\frac{\pi}{4}$.
We note that $J^{\rm dia}_{j \rightarrow i} =0$ for $|\alpha|=|\phi_n|=\frac{\pi}{2}$, which means that for these bonds ${\bm J}_{j \rightarrow i}={\bm J}^{\rm para}_{j \rightarrow i}$. 
Such ${\bm J}^{\rm para}_{j \rightarrow i}$ flowing on the 
bonds perpendicular to ${\bm A}$ 
is unique to the 
non-uniform 
superconductor, and the presence of ${\bm J}^{\rm para}_{j \rightarrow i}$ prevents the formation of 
the one-dimensional channels. 

As we show in Appendix B, results for the Penrose structure show that perpendicular currents are likewise present in that case. Indeed, perpendicular local currents can arise in quasiperiodic structures as these systems do not have translation invariance. There are no such currents on the square or honeycomb lattices (see Appendix B).

Furthermore, we have checked that such currents flow even when the local order parameter $\Delta_i$ and electron density $n_i$ in Eq.~(\ref{eq:BdG}) are assumed to be uniform on all sites (self-consistency is not imposed). This shows that the non-uniformity of $\Delta_i$ and $n_i$ are not an essential condition for perpendicular currents to flow in this case.

\begin{figure}
    \centering
    \includegraphics[width=\linewidth]{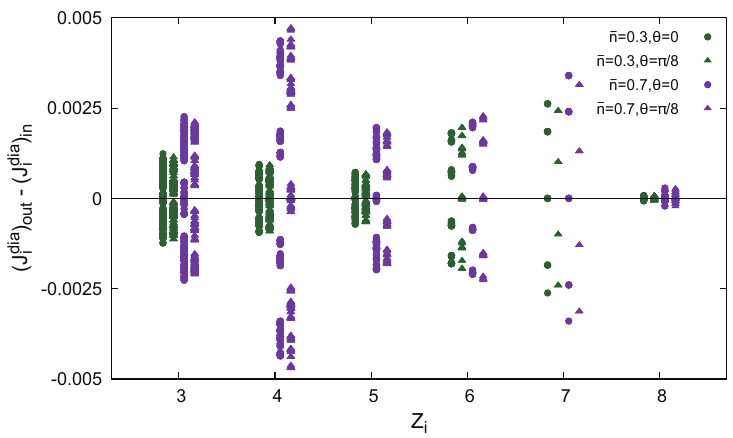}
    \caption{Divergence 
    $(J^{\rm dia}_{i})_{\rm out}
    -(J^{\rm dia}_{i})_{\rm in}$ of the diamagnetic current on the site $i$ at $T=0.01$ for (i) ${\bar n}=0.3, \theta=0$, (ii) ${\bar n}=0.3, \theta=\frac{\pi}{8}$, (iii) ${\bar n}=0.7, \theta=0$, and (iv) ${\bar n}=0.7, \theta=\frac{\pi}{8}$. The distributions are classified according to the coordination 
    number $Z_i$.
    The results for each $Z_i$ is plotted with the abscissa value shifted for each condition.}
    \label{fig:Jdia_conservation}
\end{figure}

Here, the existence of the paramagnetic current in these directions can be understood in terms of the conservation law of the local supercurrent, which is defined as that 
the total currents coming in and going out of each site should agree.
To 
see this more clearly,
in Fig.~\ref{fig:Jdia_conservation},
we show the divergence $(J^{\rm dia}_{i})_{\rm out}-(J^{\rm dia}_{i})_{\rm in}$ of $J^{\rm dia}_{j \rightarrow i}$ at each site 
with classifying the sites by the coordination number $Z_i$.
The diamagnetic currents entering and leaving a site $i$ are expressed as $(J^{\rm dia}_{i})_{\rm in}$ and $(J^{\rm dia}_{i})_{\rm out}$, respectively.
We note that the local supercurrent
${\bm J}_{j \rightarrow i}$ 
is conserved at any site in both the periodic and quasiperiodic systems. While the local current must be conserved, as required by gauge invariance, this constraint does not apply to the diamagnetic and paramagnetic parts taken separately. 
In periodic systems with uniform superconducting states, one finds nevertheless that the diamagnetic and paramagnetic 
currents are separately conserved. 
That is because they are proportional to scalar products of ${\bm A}$ and ${\bm r}_{ij}$ and the summation of the most neighbor ${\bm r}_{ij}$ is zero at all sites $i$. 
In contrast, in the quasiperiodic system, we observe that
the diamagnetic currents are not locally conserved. This can be seen from the plot in Fig.~\ref{fig:Jdia_conservation} which shows that 
the divergence of the local diamagnetic current is not zero.
This leads to the fact that the paramagnetic current is not locally conserved, either, in order to satisfy the local current conservation of ${\bm J}_{j \rightarrow i}={\bm J}^{\rm para}_{j \rightarrow i}+{\bm J}^{\rm dia}_{j \rightarrow i}$.

In addition to the $\phi_n$ dependence,
we see in Fig.~\ref{fig:xyJ}(a) a trend
that $J^{\rm dia}_{j \rightarrow i}$ becomes larger on bonds connected to the sites with larger coordination numbers such as $Z_i=8$, 7, and 6 where the local electron density $n_i$ is larger.
As expected from the physical role of ${\bm J}^{\rm para}_{j \rightarrow i}$,
it flows in the opposite direction to that of ${\bm J}^{\rm dia}_{j \rightarrow i}$.
Therefore, the paramagnetic current also flows more on the bonds connected to the sites with larger $Z_i$. 
This point will be further discussed in Sec.~\ref{sec:results}C.
Importantly, this paramagnetic current remains finite even at zero temperature as pointed out in Ref.~\cite{liu2022cooper} (see Sec.~\ref{subsec:temperature}), contrary to the case of the uniform system. 
Moreover,
${\bm J}^{\rm para}_{j \rightarrow i}$
remains finite regardless of the flow directions.

Summing up, 
we have described the spatial distribution of the local supercurrent ${\bm J}_{j \rightarrow i}$ on the Ammann-Beenker structure for $\theta=0$ (vector potential along the $x$-axis). We find that $J_{j \rightarrow i}$ is inhomogeneous, and takes different values on the tiling, 
even among the bonds sharing the same bond factor.
In contrast to the case of periodic systems, $J^{\rm dia}_{j \rightarrow i}$ itself is not locally conserved on this structure 
but is compensated by $J^{\rm para}_{j \rightarrow i}$.
One of the consequences of this type of compensation is that $J^{\rm para}_{j \rightarrow i}$ flows on the transverse bonds of $|\phi_n|=\frac{\pi}{2}$ where no diamagnetic current flows. 
We stress that this effect is observable only upon examining current patterns at a given node, that is at a local scale. 

\subsection{Filling ${\bar n}$ dependence}

\begin{figure*}[tb]

\centering
\includegraphics[width=\textwidth]{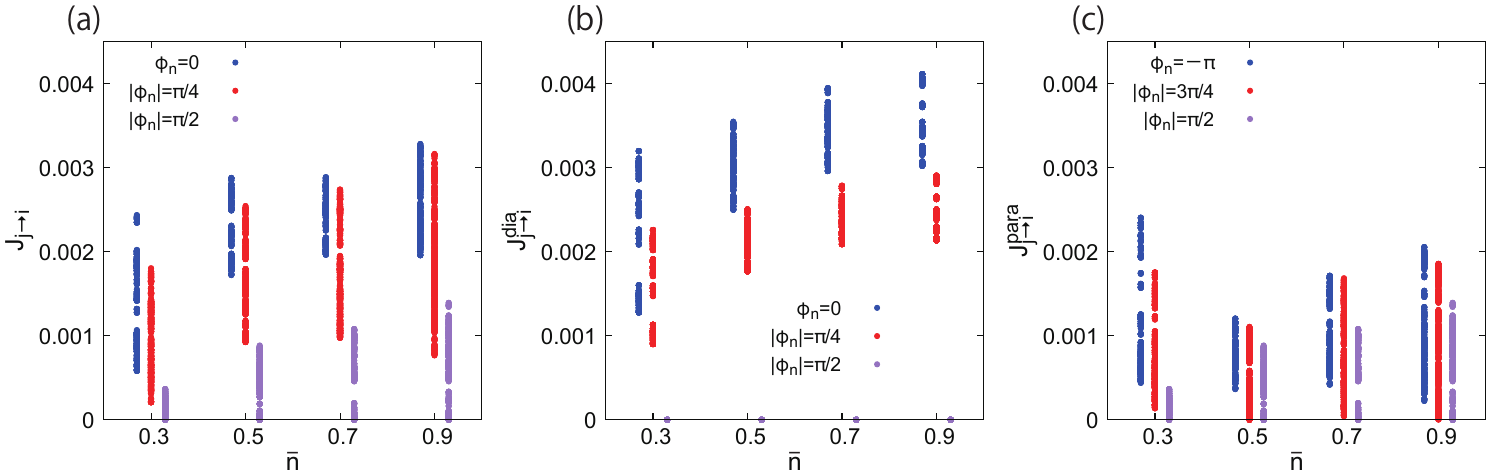}




\caption{\label{fig:Jn}
 Filling dependence of the local current $J_{j \rightarrow i}$ (a), diamagnetic current $J^{\mathrm {dia}}_{j \rightarrow i}$ (b) and paramagnetic current $J^{\mathrm {para}}_{j \rightarrow i}$ (c) for $U=-3$, $T=0.01$, and $\theta=0$. 
 The results in the panel (a) are separated into the diamagnetic and paramagnetic components respectively in the panels (b) and (c). We note that $J^{\rm para}_{j \rightarrow i}$ flows in the opposite direction of $J^{\rm dia}_{j \rightarrow i}$. The data points for each $|\phi_n|$ are slightly shifted in the horizontal direction for the sake of visibility.
 }
\end{figure*}

Since 
the distribution of $n_i$ and $\Delta_i$ changes significantly with the filling as shown in Fig.~\ref{fig:n-op}, 
the supercurrent distribution is also expected to change accordingly. 
First, we compare the spatial structure of the supercurrent for ${\bar n}=0.7$ in Fig.~\ref{fig:xyJ}(b) and ${\bar n}=0.3$ in Fig.~\ref{fig:xyJ}(a). 
In the case of ${\bar n}=0.7$, the distribution 
of ${\bm J}^{\rm dia}_{j \rightarrow i}$ 
becomes relatively uniform for the same $|\phi_n|$, and each $J^{\rm dia}_{j \rightarrow i}$ is larger than 
that for ${\bar n}=0.3$. 
At the same time, $J^{\rm para}_{j \rightarrow i}$ around the $Z_i=8$, 7, and 6 sites is strongly reduced from that for $\bar{n}=0.3$. 
Moreover, the much larger 
$J^{\rm para}_{j \rightarrow i}$ flows in the direction of $|\phi_n|=\frac{\pi}{2}$. 
We observed that this trend is particularly pronounced 
on bonds 
connected to the $Z_i=4$ sites, where $n_i$ increases significantly with $\bar{n}$ in Fig.~\ref{fig:n-op}(c).

To see the $\bar{n}$ dependence more systematically, we plot $J_{j \rightarrow i}$, $J^{\rm dia}_{j \rightarrow i}$, and $J^{\rm para}_{j \rightarrow i}$ against $\bar{n}$ in Fig.~\ref{fig:Jn}.
Since
$\phi_n$ and $-\phi_n$ are equivalent for $\theta=0$, the distributions 
are grouped by $|\phi_n|$ in Fig.~\ref{fig:Jn}.
The flow directions $\phi_n$ of ${\bm J}^{\rm para}_{j \rightarrow i}$ is rotated by $\pi$ from those of ${\bm J}^{\rm dia}_{j \rightarrow i}$ in Fig.~\ref{fig:Jn}(c) 
since it flows in the opposite direction of ${\bm J}^{\rm dia}_{j \rightarrow i}$.
In Fig.~\ref{fig:Jn}(b), we see a trend that $J^{\rm dia}_{j \rightarrow i}$ of $|\phi_n|=0$ and $\frac{\pi}{4}$ 
increases with ${\bar n}$. 
Note that $J^{\rm dia}_{j \rightarrow i}$ of $|\phi_n|=\frac{\pi}{2}$ is 0 due to the bond factor.
In Fig.~\ref{fig:Jn}(c), while we do not find a clear trend in $J^{\rm para}_{j \rightarrow i}$ of $|\phi_n|=0$ and $\frac{\pi}{4}$, we find that 
$J^{\rm para}_{j \rightarrow i}$
of $|\phi_n|=\frac{\pi}{2}$ and its distribution range  increase monotonically with $\bar n$. 
This suggests that the conservation law of the diamagnetic current is further violated as the filling ${\bar n}$ increases.
As shown in Fig.~\ref{fig:Jdia_conservation} for ${\bar n}=0.7$, the deviation from 0 
becomes big compared to the case of ${\bar n}=0.3$. 
Interestingly, the divergence at $Z_i=4$, where $n_i$ and $|\Delta_i|$ are particularly bigger than those of ${\bar n}=0.3$ [Figs.~\ref{fig:n-op}(c) and (d)], ranges much larger than that for ${\bar n}=0.3$. 
As the net result, 
reflecting the increase of $J^{\rm dia}_{j \to i}$, $J_{j \to i}$ of $|\phi_n|=0$ and $\frac{\pi}{4}$ increase with $\bar n$. 
However, 
the distribution ranges of $J_{j \to i}$ in those directions 
do not show a monotonic change with respect to $\bar{n}$ while it monotonically increases in the $|\phi_n|=\frac{\pi}{2}$ direction.
In this way, the current distribution strongly depends on $\bar n$.

\begin{figure}
    \centering
    \includegraphics[width=\linewidth]{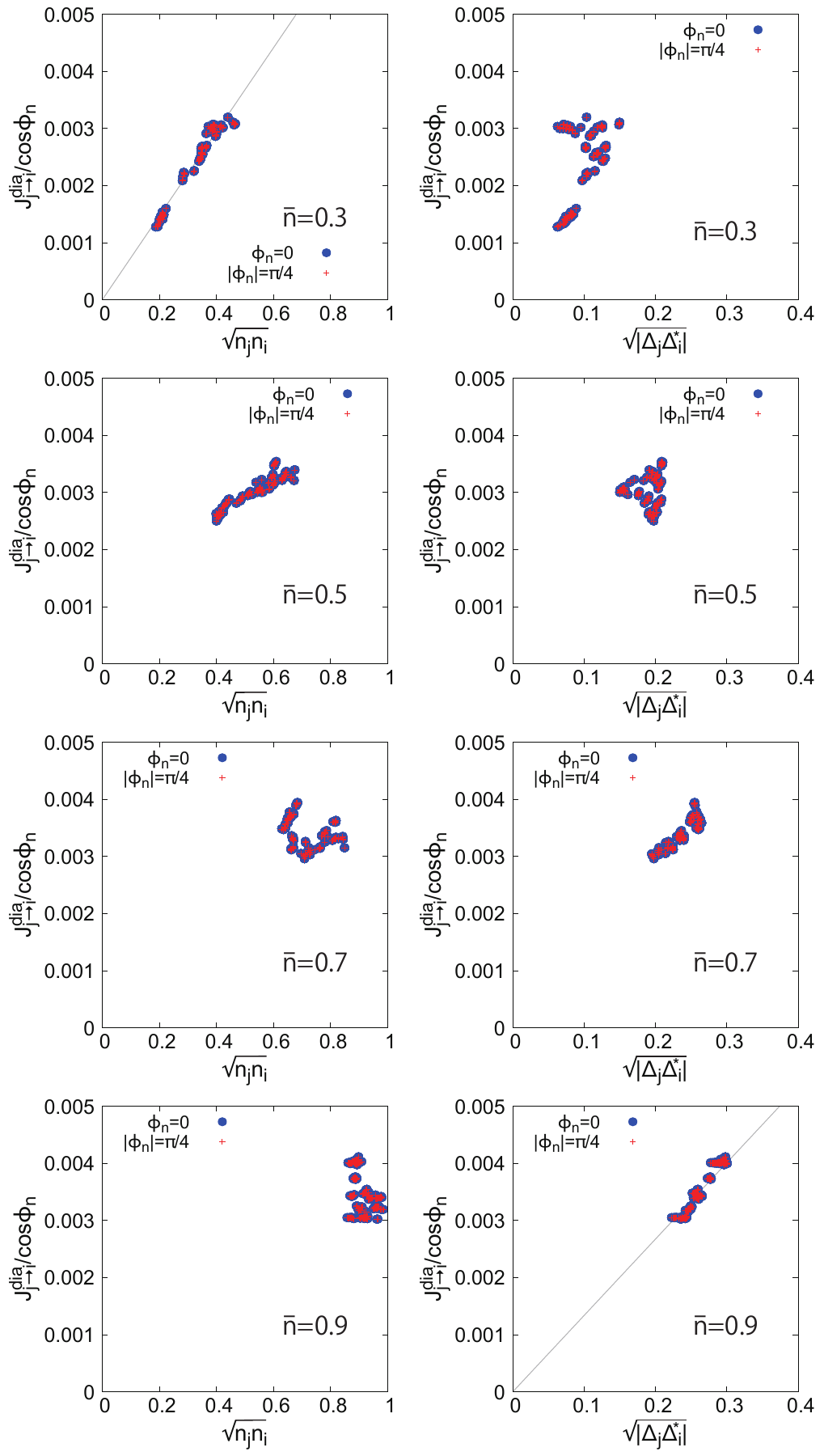}
    \caption{\label{fig:Jdia-nD}
    Values of $J^{\rm dia}_{j \rightarrow i}$ divided by the bond factor $\cos \phi_n$ are 
    plotted versus $\sqrt{n_j n_i}$ (left column) and $\sqrt{|\Delta_j \Delta^\ast_i|}$ (right column) for four different fillings $\bar n$. Parameters : $U=-3, T=0.01$, and $\theta=0$.
    } 
\end{figure}

It is interesting at this point to ask what factors determine the current distribution aside from the trivial bond factors. To study this question, we investigate the distribution $J^{\rm dia}_{j \rightarrow i}$ after dividing by the bond factor (for bond angles $|\phi_n| = 0$ and $\frac{\pi}{4}$ where $\cos\phi_n\neq0$). The relation $J^{\rm dia}_{j \rightarrow i} \sim {\rm Re}\left\{ \sum_{\sigma}\langle{\hat{c}^{\dag}_{i\sigma}\hat{c}_{j\sigma}\rangle} \right\}$ (from Eq.~(\ref{eq:Jdia})) suggests that there may exist two types of simplified dependence. The rescaled supercurrent variable is thus plotted in two different ways in Fig.~\ref{fig:Jdia-nD}: as a function of $\sqrt{n_j n_i}$ (left-hand column), and as a function of $\sqrt{|\Delta_j \Delta^\ast_i|}$ (right-hand column), for four different values of the 
filling. 
As one can see in the figure, the blue ($\phi_n=0$) and red dots ($|\phi_n|=\frac{\pi}{4}$) overlap, showing that the new variables $J^{\rm dia}_{j \rightarrow i}/\cos\phi_n$ are independent of the bond orientation. The plots show 
that systematic correlations do exist between the rescaled local currents and the local charge/order parameter in some limits.
For small filling, 
${\bar n}=0.3$ (top row), $J^{\rm dia}_{j \rightarrow i}$ is 
positively correlated with $\sqrt{n_j n_i}$, but is uncorrelated with the local superconducting order parameter amplitudes. For the  large filling ${\bar n}=0.9$ (bottom row), 
the vice versa is true: the current is correlated with $\sqrt{| \Delta_j \Delta^\ast_i |}$, but is uncorrelated with the local charges. 
Based on these numerical observations we conclude that at low filling the diamagnetic current on a given bond is approximately 
$$J^{\rm dia}_{j \rightarrow i} \propto \sqrt{n_j n_i}\cos\phi_n,$$ while at higher filling, the diamagnetic current on a bond is approximately given by
$$J^{\rm dia}_{j \rightarrow i} \propto \sqrt{| \Delta_j \Delta^\ast_i |}\cos\phi_n.$$
Intermediate behaviors can be seen for ${\bar n}=0.5$ and 0.7, showing that both amplitude and phase variations are important in the generic case.
The limiting behaviors for small and large fillings help to explain our observations: at small filling $\bar{n}$, the first relation attributes the large $J^{\rm dia}_{j \rightarrow i}$ around the sites with $Z_i=8$ and 7 
in Fig.~\ref{fig:xyJ}(a) to the 
large $n_i$ at such sites [Fig.~\ref{fig:n-op}(a)]. 
In the opposite limit of 
a high $\bar n$, the second relation accounts for the observation in Figs.~\ref{fig:xyJ}(b) and \ref{fig:n-op}(d) that $J^{\rm dia}_{j \rightarrow i}$ flows well on bonds connected to the sites with $Z_i=3$ and 4, where $|\Delta_i|$ is large. 


\subsection{Temperature $T$ dependence} 
\label{subsec:temperature}

\begin{figure*}[tb]
\centering
\includegraphics[width=\textwidth]{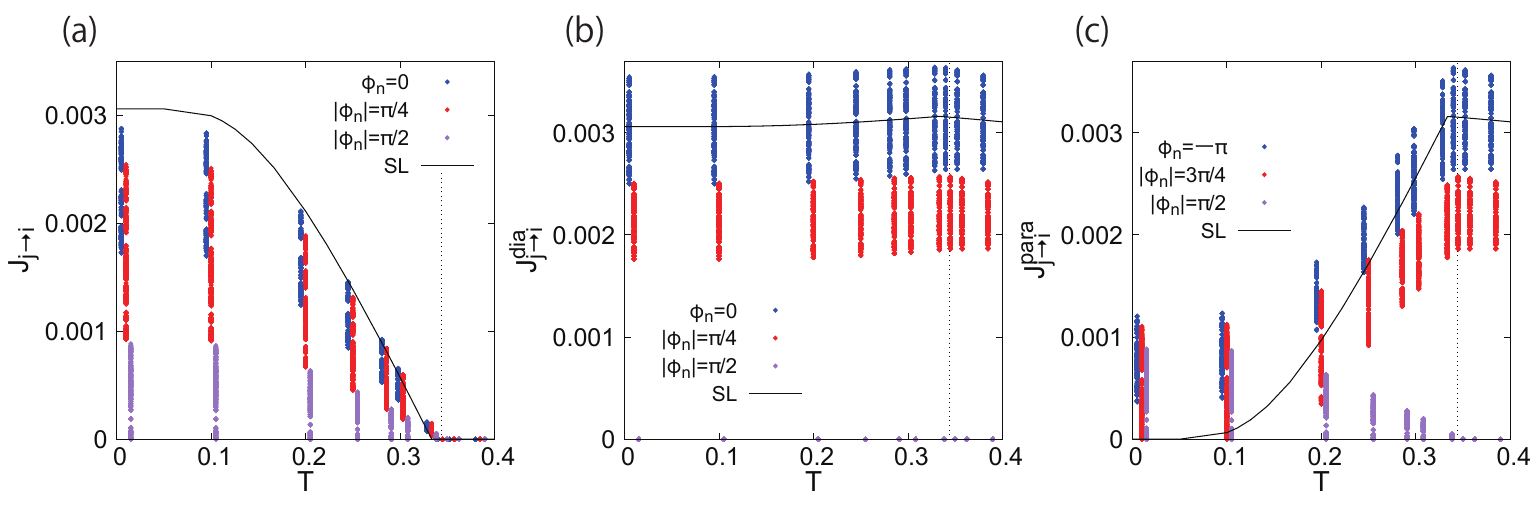}




\caption{\label{fig:TJ}
  Temperature dependence of the local current $J_{j \rightarrow i}$ (a),  diamagnetic current $J^{\rm {dia}}_{j \rightarrow i}$ (b) and paramagnetic current $J^{\mathrm {para}}_{j \rightarrow i}$ (c) for $U=-3$, ${\bar n}=0.5$, and $\theta=0$. Results on a square lattice (SL) of 900 sites 
  are shown by black 
  curves.
  The vertical dotted line represents $T_{\rm c}$ of the Ammann-Beenker structure. 
  The results in the panel (a) are separated into the diamagnetic and paramagnetic components respectively in the panels (b) and (c). We note that $J^{\rm para}_{j \rightarrow i}$ flows in the opposite direction of $J^{\rm dia}_{j \rightarrow i}$. The data points for each $|\phi_n|$ are slightly shifted in the horizontal direction for the sake of visibility.
 }
\end{figure*}



 
The superconducting transition temperature $T_{\rm c}$ for the Ammann-Beenker structure is found using the condition $|\Delta_i(T_{\rm c})| = 0$ for all the sites. For $U=-3$ and 
filling ${\bar n}=0.5$, the value of $T_{\rm c}=0.344$ for the Ammann-Beenker structure, which can be compared with a value of 0.333 for the square lattice on the same interaction strength and filling within our framework.

Figures~\ref{fig:TJ}(a), (b) and (c) show the temperature dependence of the local current and its dia- and para- components respectively.  
Black curves show 
the results for a square lattice ($N=900$) 
with the same parameters. One sees that the local supercurrent in the tiling tends to zero as $T$ approaches $T_{\rm c}$, in accordance with expectation.

Fig.~\ref{fig:TJ}(b) shows that $J^{\rm dia}_{j \rightarrow i}$ 
is almost constant as a function of temperature for all the bond orientations. 
Note that the diamagnetic current is identically 0 due to the bond factor for $|\phi_n|=\frac{\pi}{2}$.

As $J^{\rm para}_{j \rightarrow i}$ cancels $J^{\rm dia}_{j \rightarrow i}$ above $T_{\rm c}$, 
the supercurrent vanishes at $T \ge T_{\rm c}$ as shown in Fig.~\ref{fig:TJ}(c).
With lowering temperature ($T<T_{\rm c}$), $J^{\rm para}_{j \rightarrow i}$ 
of $|\phi_n|=0$ and $\frac{\pi}{4}$ decreases. 
This tendency is similar to the results on the square lattice. 
However, while the paramagnetic contribution vanishes 
at $T \rightarrow 0$ in the square lattice, it exhibits a non-zero value even at zero temperature in the quasiperiodic structure, as pointed out in Ref.~\cite{liu2022cooper}, for the site-averaged values. Our results reveal that it holds for all flow directions $\phi_n$. 
Remarkably, in the directions of $|\phi_n|=\frac{\pi}{2}$, $J^{\rm para}_{j \rightarrow i}$ increases on lowering $T$ $(<T_{\rm c})$. 

The existence of ${\bm J}^{\rm para}_{j \rightarrow i}$ even at $T \rightarrow 0$ can be qualitatively understood in terms of the 
finite center-of-mass
momentum ${\bm p}$ of the Cooper pairs.
\begin{eqnarray}
\label{eq:momentum}
    m^\ast \langle {\bm v} \rangle
    = \langle {\bm p} \rangle - e^\ast 
    \langle {\bm A} \rangle / c .
\end{eqnarray}
Here $m^\ast$, ${\bm v}$, and $e^\ast$ respectively denote the mass, velocity, and charge of the Cooper pairs, and $c$ denotes the light velocity. In the quasiperiodic systems, the Cooper pairs hold finite canonical momentum, as pointed out in Ref.~\cite{Sakai2017}. Therefore, the first term in Eq.~(\ref{eq:momentum}) does not vanish and 
gives a finite contribution 
to
${\bm J}^{\rm para}_{j \rightarrow i}$ even at $T=0$.
\subsection{Applied angle $\theta$ dependence}

\begin{figure*}[tb]

\centering
\includegraphics[width=\textwidth]{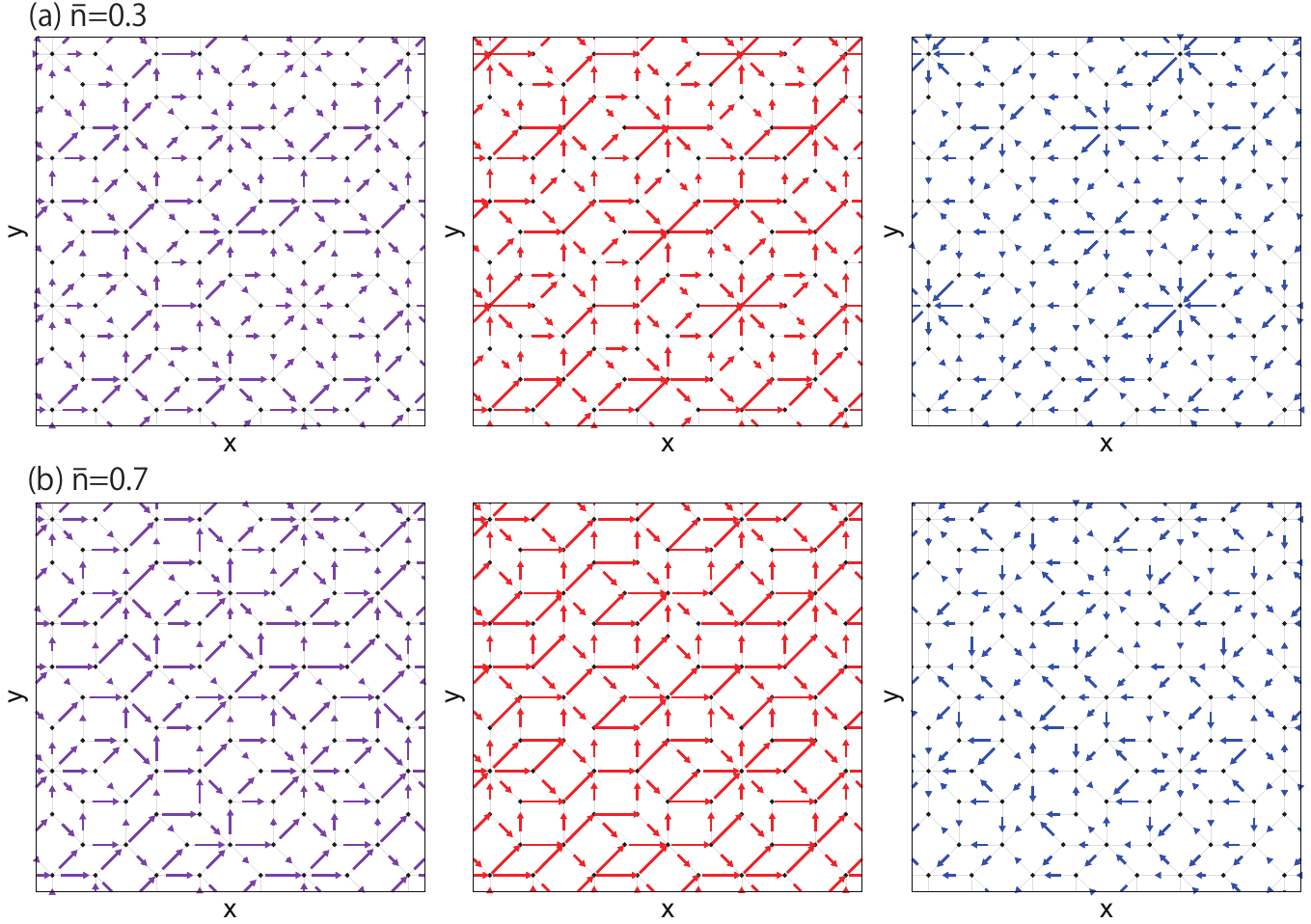}







\caption{
  \label{fig:xyJ225}
  The same as Fig.~\ref{fig:xyJ}, but for $\theta=\frac{\pi}{8}$. 
}
\end{figure*}

\begin{figure*}[tb]

\centering
\includegraphics[width=\textwidth]{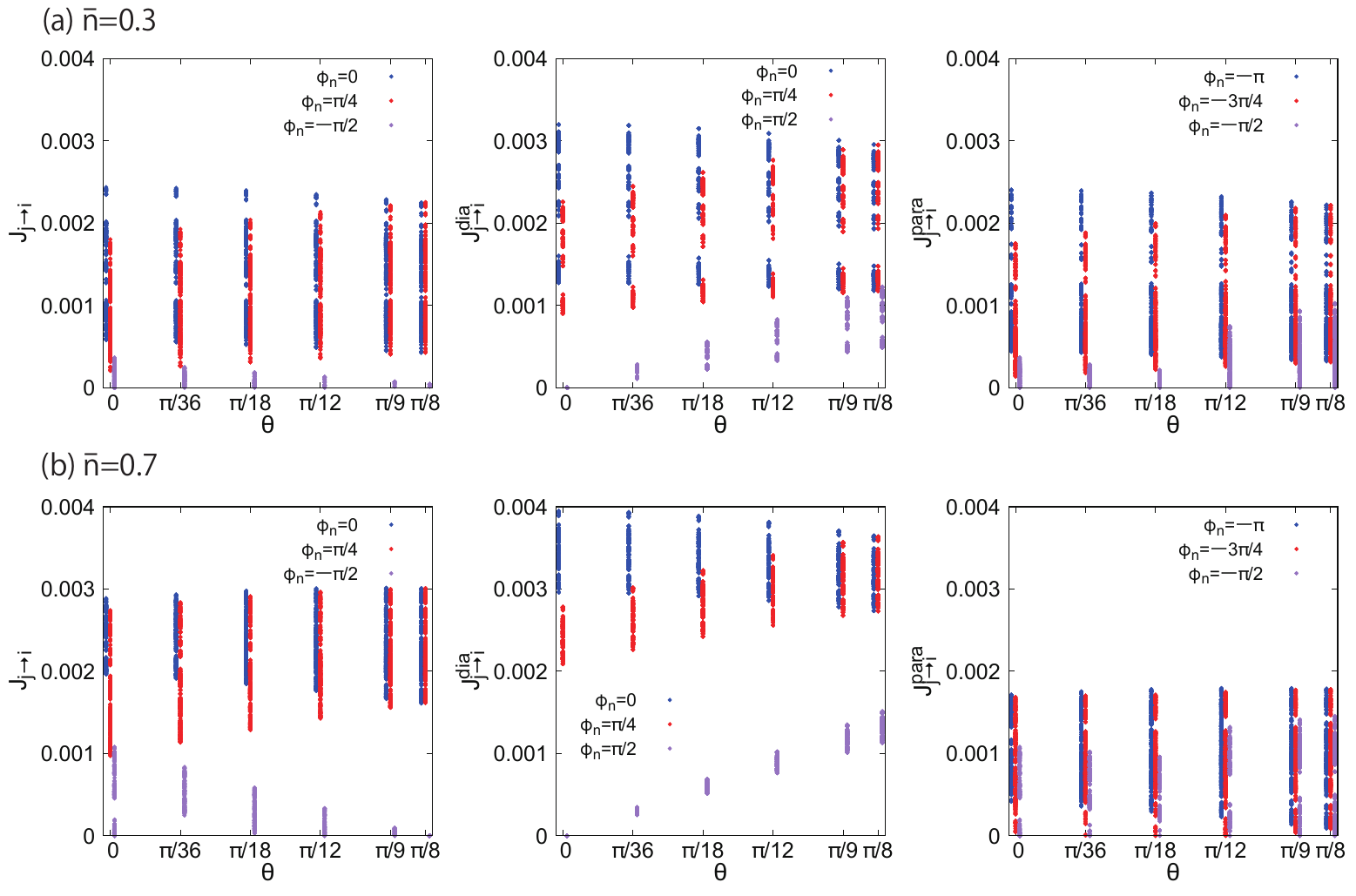}







\caption{
  \label{fig:J-theta}
  Distributions of the local current $J_{j \rightarrow i}$ (left panels), diamagnetic current $J^{\rm dia}_{j \rightarrow i}$ (middle panels) and paramagnetic current $J^{\rm para}_{j \rightarrow i}$ (right panels) for every $\frac{\pi}{36}$ [rad] from $\theta=0$ [rad] to $\frac{\pi}{8}$ [rad] at $U=-3$ and $T=0.01$. The distribution of $J_{j \rightarrow i}$ at each $\theta$ is classified 
  by the flow directions $\phi_n=0$, $\frac{\pi}{4}$, and $-\frac{\pi}{2}$ for (a) ${\bar n}=0.3$ and (b) ${\bar n}=0.7$. The results in the panel (a) are separated into the diamagnetic and paramagnetic components respectively in the panel (b) and (c). The data points for each $\phi_n$ are slightly shifted in the horizontal direction for the sake of visibility.
}
\end{figure*}

In this final section, we 
focus on the spatial distribution of the local supercurrent when
the angle $\theta$ of the applied vector potential ${\bm A}$ is varied.  
From the eight-fold symmetry of the system, one expects 
that when ${\bm A}$ is applied in the diagonal direction ($\theta = \frac{\pi}{4}$), the real-space distribution of the supercurrent should be essentially 
the same as Fig.~\ref{fig:xyJ} after $\frac{\pi}{4}$ rotation. 
(Note that in practice, the perfect 8-fold symmetry is slightly broken in 
the approximants, with additional symmetry breaking due to toroidal boundary conditions along the $x$ and $y$ directions. These effects depend on the size of the approximants, and we have checked that they are small for our system size of $N=1393$ sites.) Therefore, we 
consider the case of $\theta=\frac{\pi}{8}$ in the following.

Figures~\ref{fig:xyJ225}(a) and (b) show the spatial distributions of local supercurrent, and its dia- and para- components, for two different fillings. 
The real-space structure of the local supercurrent ${\bm J}_{j \rightarrow i}$ at $\theta=\frac{\pi}{8}$ is shown in the left panels of Figs.~\ref{fig:xyJ225}(a) and (b) in two cases of ${\bar n} = 0.3$ and 0.7. 
Since the bond factor $\cos\alpha$ has the same 
value on bonds of $\phi_n=0$ and $\frac{\pi}{4}$ 
for $\theta=\frac{\pi}{8}$, spatial structures of ${\bm J}_{j \rightarrow i}$ are 
intermediate between the flow pattern for $\theta = 0$ (Fig.~\ref{fig:xyJ}) and its $\frac{\pi}{4}$ rotation.

We now examine the diamagnetic currents 
${\bm J}^{\rm dia}_{j \rightarrow i}$,
which are shown in the middle panels. 
Note firstly that in the case of $\theta=\frac{\pi}{8}$, all bond factors are non-zero, resulting 
in non-zero ${\bm J}^{\rm dia}_{j \rightarrow i}$ for all the directions. 
For ${\bar n}=0.3$, 
one sees that ${\bm J}^{\rm dia}_{j \rightarrow i}$ 
is larger on bonds
connected to the sites with 
a larger
$Z_i$. On the other hand, for ${\bar n}=0.7$, 
such a tendency is less clear
and ${\bm J}^{\rm dia}_{j \rightarrow i}$ depends principally on the bond orientation.
These characteristics of ${\bm J}^{\rm dia}_{j \rightarrow i}$ distribution and its dependence on ${\bar n}$ resemble those for $\theta=0$ already described in Secs.~\ref{sec:results}.B and C. This is to be expected, 
since changing $\theta$ results in changing the phase of the wave  functions $u_\epsilon({\bm r}_i)$ and $v_\epsilon({\bm r}_i)$, 
but not their absolute values and hence $n_i$ and $|\Delta_i|$ given by Eqs.~(\ref{eq:gap-eq}), (\ref{eq:density1-eq}), and (\ref{eq:density2-eq}) do not change.

The right panels of Fig.~\ref{fig:xyJ225} show the spatial distribution of the paramagnetic current ${\bm J}^{\rm para}_{j \rightarrow i}$. 
Reflecting the behavior of 
${\bm J}^{\rm dia}_{j \rightarrow i}$ described 
above, 
${\bm J}^{\rm para}_{j \rightarrow i}$ flows to recover the current conservation of ${\bm J}_{j \rightarrow i}$.

The detailed $\theta$ dependence of $J^{\rm dia}_{j \rightarrow i}$ is displayed in the middle panels of Fig.~\ref{fig:J-theta}.
For $\phi_n=0$, $J^{\rm dia}_{j \rightarrow i}$ for both ${\bar n}=0.3$ and 0.7 has relatively large
values at $\theta=0$ and gradually decreases as $\theta$ increases from 0 to $\frac{\pi}{8}$.
On the other hand, in the direction of $\phi_n=\frac{\pi}{4}$, $J^{\rm dia}_{j \rightarrow i}$ increases with increasing $\theta$.
Finally, the distributions of $\phi_n=0$ and $\frac{\pi}{4}$ components coincide at $\theta=\frac{\pi}{8}$.
The changes with $\theta$ for $\phi_n=0$ and $\frac{\pi}{4}$ approximately reflect the bond factor.
The average of $J^{\rm dia}_{j \rightarrow i}$ at $\phi_n=\frac{\pi}{2}$ increases as $\theta$ increases.
In addition, we confirmed that 
the conservation law of $J^{\rm dia}_{j \rightarrow i}$ is violated
for 
$\theta=\frac{\pi}{8}$, too, as shown in Fig.~\ref{fig:Jdia_conservation}.

In the right panel of Fig.~\ref{fig:J-theta}(a) for ${\bar n}=0.3$, 
we see that $J^{\rm para}_{j \rightarrow i}$ of $\phi_n=-\pi$ ($-\frac{3\pi}{4}$) becomes smaller (bigger) as $\theta$ increases.
On the other hand, the $\theta$ dependence of $J^{\rm para}_{j \rightarrow i}$ for ${\bar n}=0.7$ is clearly weaker as shown in the right panel of Fig.~\ref{fig:J-theta}(b).
We note that $J^{\rm para}_{j \rightarrow i}$ 
in the direction of  $\phi_n=-\frac{\pi}{2}$, which is related to a back-flow as discussed later, decreases 
with $\theta$ for $0 \le \theta \le \frac{\pi}{18}$, while it increases 
for $\frac{\pi}{18} < \theta \le \frac{\pi}{8}$ to cancel the increase of $J^{\rm dia}_{j \rightarrow i}$.

The left panels of Fig.~\ref{fig:J-theta} show the distributions of the resulting $J_{j \rightarrow i}$ for various angles $\theta$.
After changes of $J_{j \rightarrow i}$ distribution with increasing $\theta$, $J_{j \rightarrow i}$ for $\phi_n=0$ and $\frac{\pi}{4}$ 
reach the same distribution at $\theta = \frac{\pi}{8}$.
In the case of ${\bar n}=0.3$, while the distribution of $J_{j \rightarrow i}$ in the direction of $\phi_n=0$ has weak $\theta$ dependence, ${\rm max}\{ J_{j \rightarrow i} \}$ decreases with $\theta$ in the range $0 \le  \theta \le \frac{\pi}{8}$. 
On the other hand, for ${\bar n}=0.7$, the distribution range of $J_{j \rightarrow i}$ for $\phi_n=0$ expands as $\theta$ increases, 
with the increase of ${\rm max}\{ J_{j \rightarrow i} \}$ 
for $0 \leq  \theta \leq \frac{\pi}{8}$.

\begin{figure}
    \centering
    \includegraphics[width=0.4\textwidth]{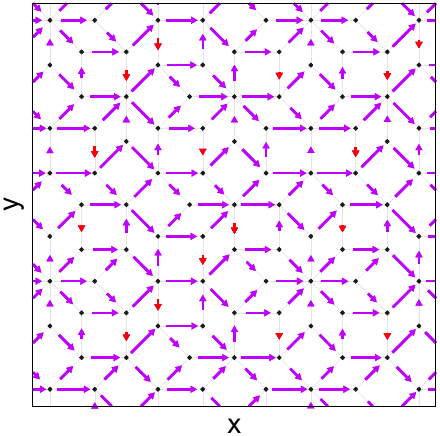}
    \caption{
    Real-space distribution of the local supercurrent ${\bm J}_{j \rightarrow i}$ for ${\bar n}=0.7$ and $\theta=\frac{\pi}{36}$. The back-flows are shown as the red downward arrows. We show a part of the system consisting of about 100 sites for visibility.
    }
    \label{fig:backflow}
\end{figure}

We have already noted that for $\theta = 0$ 
the current 
can 
flow in the direction transverse to the vector potential. For example, ${\bm J}_{j \rightarrow i}$ for $\phi_n = -\frac{\pi}{2}$ is a transverse flow to ${\bm A}$, in the results shown in Fig.~\ref{fig:xyJ}. 
When $\theta \neq 0$, 
a ``back-flow", i.e., the current satisfying
${\bm J}_{j \rightarrow i}\cdot {\bm A}
\ \propto \cos\alpha \  
<0$
, occurs. 
Figure~\ref{fig:backflow} shows
an example of the back-flow in the case of ${\bar n}=0.7$ and $\theta=\frac{\pi}{36}$.
In the left panels of Figs.~\ref{fig:J-theta}(a) and (b), except for $\theta=0$, ${\bm J}_{j \rightarrow i}$ for $\phi_n = -\frac{\pi}{2}$ is the back-flow. 
We find that 
this back-flow decreases with $\theta$ 
up to $\theta = \frac{\pi}{8}$. 
Since $J^{\rm para}_{j \rightarrow i}$ flowing in the direction of $\phi_n=-\frac{\pi}{2}$ is larger than $J^{\rm dia}_{j \rightarrow i}$ in the opposite direction, the back-flow appears 
in ${\bm J}_{j \rightarrow i}$. 

In this way, the back-flow comes from the paramagnetic component, which flows to satisfy the local current conservation of 
${\bm J}_{j \rightarrow i}$, compensating for the broken local current conservation of 
${\bm J}^{\rm dia}_{j \rightarrow i}$. 
Thus, the back-flow of ${\bm J}_{j \rightarrow i}$ is also one of the 
characteristics of the quasiperiodic 
superconductors.

\section{Summary}
\label{sec:sum}

We have studied the local supercurrent flow under the uniform vector potential on the Ammann-Beenker structure. To address this problem, we introduced the attractive Hubbard model, 
where the effect of vector potential is incorporated as the Peierls phase in the transfer term, and numerically analyzed it based on the self-consistent BdG mean-field theory. 
We decomposed the local supercurrent into the diamagnetic and paramagnetic current in our formulation in order to better understand the non-uniform spatial distribution. Our formulation for the local supercurrent is applicable not only to other 
quasiperiodic structures but also to general non-uniform 
structures with the periodic boundary condition.

We confirmed that the local electron density and superconducting order parameter are distributed non-uniformly with approximate 8-fold symmetry, as known in Refs.~\cite{Sakai2017,Takemorisc,A-Bsuper}. 
The distributions greatly 
vary depending on the 
filling ${\bar n}$. 
We 
clarified a spatial distribution of the supercurrent and its variation depending on (i) the average electron filling
 $\bar n$, (ii) temperature $T$, and (iii) 
the angle $\theta$ of the applied vector potential. 

Firstly, the diamagnetic current has 
a temperature dependence similar to that
in the uniform systems, but the paramagnetic current has a finite value even at $T\rightarrow0$ for all flow 
directions $\phi_n$. We believe that such a phenomenon can be also realized in the 
Fulde-Ferrell-Larkin-Ovchinnikov (FFLO) states~\cite{FFLO1,FFLO2}. However, in the quasiperiodic systems, proper adjustment of the magnetic field is unnecessary, and it would be 
easier to confirm this phenomenon 
through experiments. 
Secondly, as the filling increases, the vertical paramagnetic current increases, which is accompanied by a change in the distribution of the local electron density and superconducting order parameter. 
Lastly, 
the local supercurrent 
flows even in 
the direction transverse to the applied vector potential. 
Furthermore, as the angle of the vector potential increases, back-flows are observed where the bond factor $\cos\alpha$ is negative.
In any case, the paramagnetic current 
behaves differently from that
in periodic systems. This is because the diamagnetic current is affected by the distributions of the local electron density and the superconducting order parameter, and 
is
not conserved locally. As a result, an excess amount of paramagnetic current 
has
to be induced even at zero temperature as a counterpart to recover the local current conservation 
and contributes to characteristic local supercurrent behaviors on the quasiperiodic structure. 

In conclusion, we have presented the first theoretical investigation of real space distributions of the supercurrent in a structure that does not possess translation invariance but is perfectly ordered.  
The novel spatial distributions of the supercurrent 
revealed in this study are the first step in understanding the response to magnetic fields and the Meissner effect in 
quasicrystalline superconductors.
More detailed investigations of the distribution of screening currents under external fields are planned for future work. 



\begin{acknowledgments}
N. T. is supported by JSPS KAKENHI Grant No. JP16H07447, JP19H05817, and JP19H05820.
S. S. is supported by JSPS KAKENHI Grant No. JP22H04603.
M. I. is supported by JSPS KAKENHI Grant No. JP21K03471 and JP19H00657.
\end{acknowledgments}

\section{Appendix}
\subsection{Gauge transformation of Hamiltonian}
\label{sec:apxA}


The formulation of the local supercurrent in Sec.~\ref{sec:method} differs from our previous formulation~\cite{TF}. Here we will see that these two formulations are equivalent through gauge transformations.

In Eq.~(\ref{eq:hamiltonian}), the following gauge transformation is applied to the creation and annihilation operators.

\begin{eqnarray}
\label{eq:gaugetrans}  
      \hat{\tilde c}_{i \sigma} = {\hat c}_{i \sigma}{\rm e}^{-{\rm i}{\bm A} \cdot {\bm r}_i}, \quad
      \hat{\tilde c}^\dag_{i \sigma} = {\hat c}^\dag_{i \sigma}{\rm e}^{{\rm i}{\bm A} \cdot {\bm r}_i}.
\end{eqnarray}
In this case, the wave functions $u_{\epsilon}({\bm r}_i), v_{\epsilon}({\bm r}_i)$ after the gauge transformation~(\ref{eq:gaugetrans}) become
\begin{eqnarray}
\label{eq:wfgauge}
     &{\tilde u}_{\epsilon}({\bm r}_i) = u_{\epsilon}({\bm r}_i){\rm e}^{{\rm i}{\bm A} \cdot {\bm r}_i}, \quad
     {\tilde v}_{\epsilon}({\bm r}_i) = v_{\epsilon}({\bm r}_i){\rm e}^{-{\rm i}{\bm A} \cdot {\bm r}_i}.
\end{eqnarray}
The superconducting order parameter is accordingly 
\begin{eqnarray}
\label{eq:dtrans}
     \tilde{\Delta}_i = \Delta_i {\rm e}^{2{\rm i}{\bm A} \cdot {\bm r}_i}.
\end{eqnarray}
The original wave functions and superconducting order parameters have a unit cell periodicity. Therefore, the above transformation adds an extra phase factor to the transformed wave functions and superconducting order parameters under the translation operation between unit cells.

In Eq. (\ref{eq:Hmf-1}), the BdG Hamiltonian is transformed as

\begin{eqnarray} 
\label{eq:Hmf-trans1}
{\hat{\cal H}} 
 = \sum_{i,j}
\left(\begin{array}{cc}
   {\hat {\tilde c}}^\dagger_{i\uparrow} & 
   {\hat {\tilde c}}_{i\downarrow}\end{array}\right)
{\hat{\tilde{\cal H}}}_{i,j}
\left(\begin{array}{c}
   {\hat {\tilde c}}_{j\uparrow}\\
   {\hat {\tilde c}}^{\dag}_{j\downarrow}\end{array}\right),
\end{eqnarray}

with 

\begin{eqnarray} 
\label{eq:Hmf-trans2}
{\hat{\tilde{\cal H}}}_{i,j}
=\left( \begin{array}{cc} 
     {\tilde K}_{\uparrow i,j }  & {\tilde \Delta}_i \delta_{i,j} \\
     {\tilde \Delta}_i^\ast \delta_{i,j} &  - {\tilde K}^\ast_{ \downarrow i,j} 
\end{array} \right),
\end{eqnarray}
where ${\tilde K}_{\sigma i,j}=-t\delta_{\langle i,j \rangle}+(Un_{i\bar\sigma}-\mu)\delta_{i,j}$.

Considering the expressions of the supercurrent, the 
expectation values of the operators in Eqs. (\ref{eq:Jtot2})$\sim$(\ref{eq:Jdia}) are 
given by
\begin{eqnarray}
     \langle {\hat c}^\dag_{i \uparrow} {\hat c}_{j \uparrow} \rangle = \langle {\hat{\tilde c}}^\dag_{i \uparrow} {\hat {\tilde c}}_{j \uparrow} \rangle {\rm e}^{{\rm i}{\bm A} \cdot {\bm r}_{ij}}, \\
     \langle {\hat c}^\dag_{i \downarrow} {\hat c}_{j \downarrow} \rangle = \langle {\hat{\tilde c}}^\dag_{i \downarrow} {\hat{\tilde c}}_{j \downarrow} \rangle {\rm e}^{{\rm i}{\bm A} \cdot {\bm r}_{ij}}.
\end{eqnarray}
With these transformations, the local supercurrent ${\bm J}_{j \rightarrow i}$ becomes
\begin{eqnarray}
\label{Jtot3}
     {\bm J}_{j \rightarrow i} 
     = 2t {\rm Im} \left( \sum_\sigma \langle {\hat {\tilde c}}^\dag_{i \sigma} {\hat {\tilde c}}_{j \sigma} \rangle \right) {\bm r}_{ij}.
\end{eqnarray}
By separating the paramagnetic and diamagnetic current from Eq. (\ref{Jtot3}), the following expressions are obtained.
\begin{eqnarray}
{\bm J}^{\mathrm{para}}_{j \rightarrow i}
= 2t\cos \left({\bm A}\cdot {\bm r}_{ij} \right)\mathrm {Im}\left\{ {\rm e}^{{\rm i}{\bm A} \cdot {\bm r}_{ij} }\sum_{\sigma}\langle{\hat{{\tilde c}}^{\dag}_{i\sigma}\hat{{\tilde c}}_{j\sigma}\rangle} \right\}
{\bm r}_{ij},
\quad 
\nonumber \\ && 
\end{eqnarray}
\begin{eqnarray}
\label{eq:Jdia2}
{\bm J}^{\mathrm{dia}}_{j \rightarrow i}
=-2t \sin \left({\bm A}\cdot {\bm r}_{ij} \right)
\mathrm {Re}\left\{ {\rm e}^{{\rm i}{\bm A} \cdot {\bm r}_{ij}} \sum_{\sigma}\langle{\hat{{\tilde c}}^{\dag}_{i\sigma}\hat{{\tilde c}}_{j\sigma}\rangle} \right\} 
        {\bm r}_{ij}.
        \quad 
        \nonumber \\ && 
\end{eqnarray}

Thus, we see that the present formulation and that of the previous one coincide through the gauge transformations.
Our intention behind the formulation of this study is to decompose the supercurrent into paramagnetic and diamagnetic components.

\subsection{Supercurrents on the Penrose structure and honeycomb lattice}
\label{apxB}
In this section, we show results for the Penrose structure, and for a simple periodic structure -- the honeycomb lattice. These examples help to clarify the reasons for the existence of non-zero perpendicular local currents which we have reported in our paper. We would like to thank one of the referees for suggesting these calculations.

\begin{figure*}[tb]
    \centering
    \includegraphics[width=\textwidth]{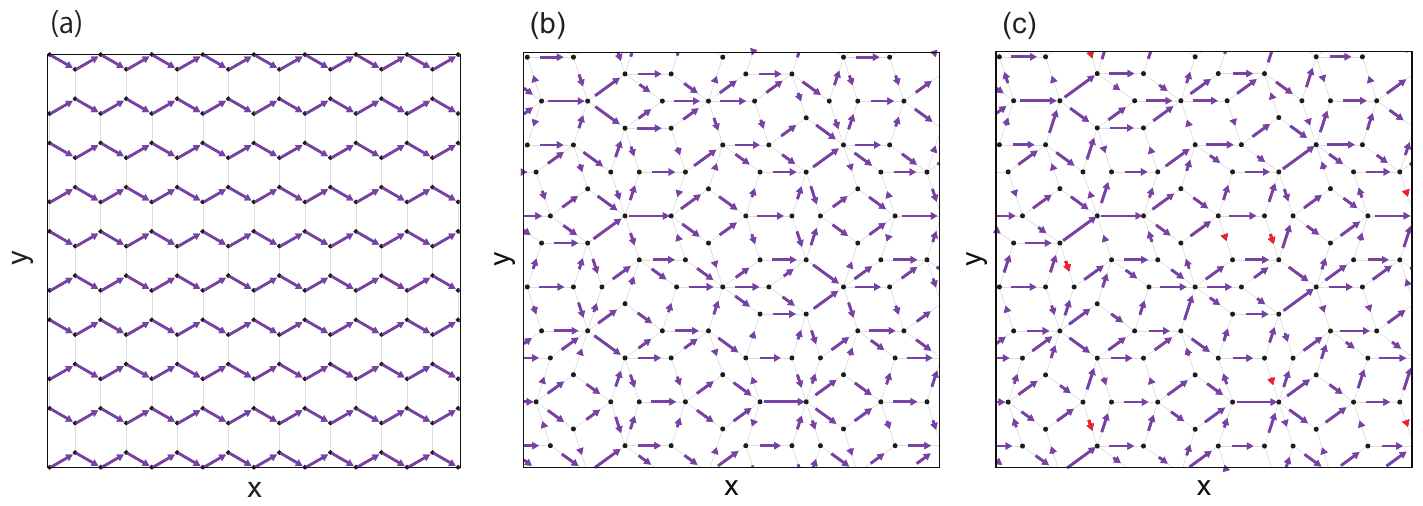}
    \caption{
    Real-space distribution of the local supercurrent ${\bm J}_{j \rightarrow i}$ on the honeycomb lattice (a) and the Penrose structure (b) and (c) at $U=-3$, $T=0.01$, ${\bar n}=0.3$. The panels (a) and (b) are for the case of $\theta=0$ while the panel (c) is in the case of $\theta=\pi/10$. The red arrows in the panel (c) show the vertical supercurrents against the applied vector potential. In each panel, we show a part of the system consisting of about 100 sites for visibility.}
    \label{fig:xy_ph}
\end{figure*}

To clarify the role played by structure and the difference in the supercurrent distribution between periodic and quasiperiodic systems, we consider the honeycomb lattice ($N=680$) and the Penrose structure ($N=644$) under periodic boundary conditions. Figure~\ref{fig:xy_ph} shows the spatial distribution of ${\bm J}_{j \rightarrow i}$ in these structures. Figures~\ref{fig:xy_ph}(a) and (b) show the case of ${\bar n}=0.3$ and $\theta=0$ on the honeycomb lattice and the Penrose structure, respectively. In the panel (a), one sees that currents are uniformly distributed along the zig-zag lines running parallel to the applied potential. One sees that there is no ${\bm J}_{j \rightarrow i}$ in the vertical direction with respect to the applied vector potential. This is expected, due to the translational and inversion symmetries of the honeycomb lattice. 
As shown in the panel (b), currents ${\bm J}_{j \rightarrow i}$ flow non-uniformly on the Penrose structure.  Figure~\ref{fig:xy_ph}(c) shows current distribution in the Penrose structure for the case of ${\bar n}=0.3$ and $\theta=\pi/10$. Flows which are perpendicular to the applied vector potential are shown in red.  
Such perpendicular currents are thus observed in both the Ammann-Beenker and Penrose structures. The above results show that while perpendicular currents are absent in simple periodic systems such as the square or the honeycomb lattices, they can exist in quasiperiodic structures.
We note, finally, that for the honeycomb lattice, $J^{\rm para}_{j \rightarrow i}$ becomes zero at $T=0$, as seen already for the square lattice. As noted in the main text, the existence of non-zero paramagnetic currents at $T=0$ is another important qualitative difference between periodic and quasiperiodic systems.


\bibliography{ref}

\begin{thebibliography}{58}%
\makeatletter
\providecommand \@ifxundefined [1]{%
 \@ifx{#1\undefined}
}%
\providecommand \@ifnum [1]{%
 \ifnum #1\expandafter \@firstoftwo
 \else \expandafter \@secondoftwo
 \fi
}%
\providecommand \@ifx [1]{%
 \ifx #1\expandafter \@firstoftwo
 \else \expandafter \@secondoftwo
 \fi
}%
\providecommand \natexlab [1]{#1}%
\providecommand \enquote  [1]{``#1''}%
\providecommand \bibnamefont  [1]{#1}%
\providecommand \bibfnamefont [1]{#1}%
\providecommand \citenamefont [1]{#1}%
\providecommand \href@noop [0]{\@secondoftwo}%
\providecommand \href [0]{\begingroup \@sanitize@url \@href}%
\providecommand \@href[1]{\@@startlink{#1}\@@href}%
\providecommand \@@href[1]{\endgroup#1\@@endlink}%
\providecommand \@sanitize@url [0]{\catcode `\\12\catcode `\$12\catcode
  `\&12\catcode `\#12\catcode `\^12\catcode `\_12\catcode `\%12\relax}%
\providecommand \@@startlink[1]{}%
\providecommand \@@endlink[0]{}%
\providecommand \url  [0]{\begingroup\@sanitize@url \@url }%
\providecommand \@url [1]{\endgroup\@href {#1}{\urlprefix }}%
\providecommand \urlprefix  [0]{URL }%
\providecommand \Eprint [0]{\href }%
\providecommand \doibase [0]{https://doi.org/}%
\providecommand \selectlanguage [0]{\@gobble}%
\providecommand \bibinfo  [0]{\@secondoftwo}%
\providecommand \bibfield  [0]{\@secondoftwo}%
\providecommand \translation [1]{[#1]}%
\providecommand \BibitemOpen [0]{}%
\providecommand \bibitemStop [0]{}%
\providecommand \bibitemNoStop [0]{.\EOS\space}%
\providecommand \EOS [0]{\spacefactor3000\relax}%
\providecommand \BibitemShut  [1]{\csname bibitem#1\endcsname}%
\let\auto@bib@innerbib\@empty
\bibitem [{\citenamefont {Shechtman}\ \emph {et~al.}(1984)\citenamefont
  {Shechtman}, \citenamefont {Blech}, \citenamefont {Gratias},\ and\
  \citenamefont {Cahn}}]{PhysRevLett.53.1951}%
  \BibitemOpen
  \bibfield  {author} {\bibinfo {author} {\bibfnamefont {D.}~\bibnamefont
  {Shechtman}}, \bibinfo {author} {\bibfnamefont {I.}~\bibnamefont {Blech}},
  \bibinfo {author} {\bibfnamefont {D.}~\bibnamefont {Gratias}},\ and\ \bibinfo
  {author} {\bibfnamefont {J.~W.}\ \bibnamefont {Cahn}},\ }\bibfield  {title}
  {\bibinfo {title} {Metallic {P}hase with {L}ong-{R}ange {O}rientational
  {O}rder and {N}o {T}ranslational {S}ymmetry},\ }\href
  {https://doi.org/10.1103/PhysRevLett.53.1951} {\bibfield  {journal} {\bibinfo
   {journal} {Phys. Rev. Lett.}\ }\textbf {\bibinfo {volume} {53}},\ \bibinfo
  {pages} {1951} (\bibinfo {year} {1984})}\BibitemShut {NoStop}%
\bibitem [{\citenamefont {Levine}\ and\ \citenamefont
  {Steinhardt}(1984)}]{qcs}%
  \BibitemOpen
  \bibfield  {author} {\bibinfo {author} {\bibfnamefont {D.}~\bibnamefont
  {Levine}}\ and\ \bibinfo {author} {\bibfnamefont {P.~J.}\ \bibnamefont
  {Steinhardt}},\ }\bibfield  {title} {\bibinfo {title} {Quasicrystals: {A}
  {N}ew {C}lass of {O}rdered {S}tructures},\ }\href@noop {} {\bibfield
  {journal} {\bibinfo  {journal} {Phys. Rev. Lett.}\ }\textbf {\bibinfo
  {volume} {53}},\ \bibinfo {pages} {2477} (\bibinfo {year}
  {1984})}\BibitemShut {NoStop}%
\bibitem [{\citenamefont {Kohmoto}\ \emph {et~al.}(1983)\citenamefont
  {Kohmoto}, \citenamefont {Kadanoff},\ and\ \citenamefont
  {Tang}}]{Kohmoto1983}%
  \BibitemOpen
  \bibfield  {author} {\bibinfo {author} {\bibfnamefont {M.}~\bibnamefont
  {Kohmoto}}, \bibinfo {author} {\bibfnamefont {L.~P.}\ \bibnamefont
  {Kadanoff}},\ and\ \bibinfo {author} {\bibfnamefont {C.}~\bibnamefont
  {Tang}},\ }\bibfield  {title} {\bibinfo {title} {Localization {P}roblem in
  {O}ne {D}imension: {M}apping and {E}scape},\ }\href
  {https://doi.org/10.1103/PhysRevLett.50.1870} {\bibfield  {journal} {\bibinfo
   {journal} {Phys. Rev. Lett.}\ }\textbf {\bibinfo {volume} {50}},\ \bibinfo
  {pages} {1870} (\bibinfo {year} {1983})}\BibitemShut {NoStop}%
\bibitem [{\citenamefont {Ostlund}\ \emph {et~al.}(1983)\citenamefont
  {Ostlund}, \citenamefont {Pandit}, \citenamefont {Rand}, \citenamefont
  {Schellnhuber},\ and\ \citenamefont {Siggia}}]{ostlund1983}%
  \BibitemOpen
  \bibfield  {author} {\bibinfo {author} {\bibfnamefont {S.}~\bibnamefont
  {Ostlund}}, \bibinfo {author} {\bibfnamefont {R.}~\bibnamefont {Pandit}},
  \bibinfo {author} {\bibfnamefont {D.}~\bibnamefont {Rand}}, \bibinfo {author}
  {\bibfnamefont {H.~J.}\ \bibnamefont {Schellnhuber}},\ and\ \bibinfo {author}
  {\bibfnamefont {E.~D.}\ \bibnamefont {Siggia}},\ }\bibfield  {title}
  {\bibinfo {title} {One-{D}imensional {S}chr{\"o}dinger {E}quation with an
  {A}lmost {P}eriodic {P}otential},\ }\href@noop {} {\bibfield  {journal}
  {\bibinfo  {journal} {Phys. Rev. Lett.}\ }\textbf {\bibinfo {volume} {50}},\
  \bibinfo {pages} {1873} (\bibinfo {year} {1983})}\BibitemShut {NoStop}%
\bibitem [{\citenamefont {Niu}\ and\ \citenamefont {Nori}(1986)}]{Qian1986}%
  \BibitemOpen
  \bibfield  {author} {\bibinfo {author} {\bibfnamefont {Q.}~\bibnamefont
  {Niu}}\ and\ \bibinfo {author} {\bibfnamefont {F.}~\bibnamefont {Nori}},\
  }\bibfield  {title} {\bibinfo {title} {Renormalization-{G}roup {S}tudy of
  {O}ne-{D}imensional {Q}uasiperiodic {S}ystems},\ }\href
  {https://doi.org/10.1103/PhysRevLett.57.2057} {\bibfield  {journal} {\bibinfo
   {journal} {Phys. Rev. Lett.}\ }\textbf {\bibinfo {volume} {57}},\ \bibinfo
  {pages} {2057} (\bibinfo {year} {1986})}\BibitemShut {NoStop}%
\bibitem [{\citenamefont {Tsunetsugu}\ \emph {et~al.}(1986)\citenamefont
  {Tsunetsugu}, \citenamefont {Fujiwara}, \citenamefont {Ueda},\ and\
  \citenamefont {Tokihiro}}]{tsunetsugu1986}%
  \BibitemOpen
  \bibfield  {author} {\bibinfo {author} {\bibfnamefont {H.}~\bibnamefont
  {Tsunetsugu}}, \bibinfo {author} {\bibfnamefont {T.}~\bibnamefont
  {Fujiwara}}, \bibinfo {author} {\bibfnamefont {K.}~\bibnamefont {Ueda}},\
  and\ \bibinfo {author} {\bibfnamefont {T.}~\bibnamefont {Tokihiro}},\
  }\bibfield  {title} {\bibinfo {title} {Eigenstates in 2-{D}imensional
  {P}enrose {T}iling},\ }\href@noop {} {\bibfield  {journal} {\bibinfo
  {journal} {J. Phys. Soc. Japan}\ }\textbf {\bibinfo {volume} {55}},\ \bibinfo
  {pages} {1420} (\bibinfo {year} {1986})}\BibitemShut {NoStop}%
\bibitem [{\citenamefont {Kohmoto}\ \emph {et~al.}(1987)\citenamefont
  {Kohmoto}, \citenamefont {Sutherland},\ and\ \citenamefont
  {Tang}}]{critical}%
  \BibitemOpen
  \bibfield  {author} {\bibinfo {author} {\bibfnamefont {M.}~\bibnamefont
  {Kohmoto}}, \bibinfo {author} {\bibfnamefont {B.}~\bibnamefont
  {Sutherland}},\ and\ \bibinfo {author} {\bibfnamefont {C.}~\bibnamefont
  {Tang}},\ }\bibfield  {title} {\bibinfo {title} {Critical wave functions and
  a {C}antor-set spectrum of a one-dimensional quasicrystal model},\
  }\href@noop {} {\bibfield  {journal} {\bibinfo  {journal} {Phys. Rev. B}\
  }\textbf {\bibinfo {volume} {35}},\ \bibinfo {pages} {1020} (\bibinfo {year}
  {1987})}\BibitemShut {NoStop}%
\bibitem [{\citenamefont {Tokihiro}\ \emph {et~al.}(1988)\citenamefont
  {Tokihiro}, \citenamefont {Fujiwara},\ and\ \citenamefont {Arai}}]{Tokihiro}%
  \BibitemOpen
  \bibfield  {author} {\bibinfo {author} {\bibfnamefont {T.}~\bibnamefont
  {Tokihiro}}, \bibinfo {author} {\bibfnamefont {T.}~\bibnamefont {Fujiwara}},\
  and\ \bibinfo {author} {\bibfnamefont {M.}~\bibnamefont {Arai}},\ }\bibfield
  {title} {\bibinfo {title} {Exact eigenstates on a two-dimensional {P}enrose
  lattice and their fractal dimensions},\ }\href
  {https://doi.org/10.1103/PhysRevB.38.5981} {\bibfield  {journal} {\bibinfo
  {journal} {Phys. Rev. B}\ }\textbf {\bibinfo {volume} {38}},\ \bibinfo
  {pages} {5981} (\bibinfo {year} {1988})}\BibitemShut {NoStop}%
\bibitem [{\citenamefont {Tsunetsugu}\ \emph {et~al.}(1991)\citenamefont
  {Tsunetsugu}, \citenamefont {Fujiwara}, \citenamefont {Ueda},\ and\
  \citenamefont {Tokihiro}}]{tsunetsugu1}%
  \BibitemOpen
  \bibfield  {author} {\bibinfo {author} {\bibfnamefont {H.}~\bibnamefont
  {Tsunetsugu}}, \bibinfo {author} {\bibfnamefont {T.}~\bibnamefont
  {Fujiwara}}, \bibinfo {author} {\bibfnamefont {K.}~\bibnamefont {Ueda}},\
  and\ \bibinfo {author} {\bibfnamefont {T.}~\bibnamefont {Tokihiro}},\
  }\bibfield  {title} {\bibinfo {title} {Electronic properties of the {P}enrose
  lattice. i. {E}nergy spectrum and wave functions},\ }\href
  {https://doi.org/10.1103/PhysRevB.43.8879} {\bibfield  {journal} {\bibinfo
  {journal} {Phys. Rev. B}\ }\textbf {\bibinfo {volume} {43}},\ \bibinfo
  {pages} {8879} (\bibinfo {year} {1991})}\BibitemShut {NoStop}%
\bibitem [{\citenamefont {Kamiya}\ \emph {et~al.}(2018)\citenamefont {Kamiya},
  \citenamefont {Takeuchi}, \citenamefont {Kabeya}, \citenamefont {Wada},
  \citenamefont {Ishimasa}, \citenamefont {Ochiai}, \citenamefont {Deguchi},
  \citenamefont {Imura},\ and\ \citenamefont {Sato}}]{Kamiya2018}%
  \BibitemOpen
  \bibfield  {author} {\bibinfo {author} {\bibfnamefont {K.}~\bibnamefont
  {Kamiya}}, \bibinfo {author} {\bibfnamefont {T.}~\bibnamefont {Takeuchi}},
  \bibinfo {author} {\bibfnamefont {N.}~\bibnamefont {Kabeya}}, \bibinfo
  {author} {\bibfnamefont {N.}~\bibnamefont {Wada}}, \bibinfo {author}
  {\bibfnamefont {T.}~\bibnamefont {Ishimasa}}, \bibinfo {author}
  {\bibfnamefont {A.}~\bibnamefont {Ochiai}}, \bibinfo {author} {\bibfnamefont
  {K.}~\bibnamefont {Deguchi}}, \bibinfo {author} {\bibfnamefont
  {K.}~\bibnamefont {Imura}},\ and\ \bibinfo {author} {\bibfnamefont {N.~K.}\
  \bibnamefont {Sato}},\ }\bibfield  {title} {\bibinfo {title} {{Discovery of
  superconductivity in quasicrystal}},\ }\href
  {https://doi.org/10.1038/s41467-017-02667-x} {\bibfield  {journal} {\bibinfo
  {journal} {Nat. Comm.}\ }\textbf {\bibinfo {volume} {9}},\ \bibinfo {pages}
  {154} (\bibinfo {year} {2018})}\BibitemShut {NoStop}%
\bibitem [{\citenamefont {Tokumoto}\ \emph {et~al.}(2023)\citenamefont
  {Tokumoto}, \citenamefont {Hamano}, \citenamefont {Nakagawa}, \citenamefont
  {Kamimura}, \citenamefont {Suzuki}, \citenamefont {Tamura},\ and\
  \citenamefont {Edagawa}}]{tokumoto23}%
  \BibitemOpen
  \bibfield  {author} {\bibinfo {author} {\bibfnamefont {Y.}~\bibnamefont
  {Tokumoto}}, \bibinfo {author} {\bibfnamefont {K.}~\bibnamefont {Hamano}},
  \bibinfo {author} {\bibfnamefont {S.}~\bibnamefont {Nakagawa}}, \bibinfo
  {author} {\bibfnamefont {Y.}~\bibnamefont {Kamimura}}, \bibinfo {author}
  {\bibfnamefont {S.}~\bibnamefont {Suzuki}}, \bibinfo {author} {\bibfnamefont
  {R.}~\bibnamefont {Tamura}},\ and\ \bibinfo {author} {\bibfnamefont
  {K.}~\bibnamefont {Edagawa}},\ }\href@noop {} {\bibinfo {title}
  {Superconductivity in a van der {W}aals layered quasicrystal}} (\bibinfo
  {year} {2023}),\ \Eprint {https://arxiv.org/abs/2307.10679} {arXiv:2307.10679
  [cond-mat.mtrl-sci]} \BibitemShut {NoStop}%
\bibitem [{\citenamefont {Bardeen}\ \emph {et~al.}(1957)\citenamefont
  {Bardeen}, \citenamefont {Cooper},\ and\ \citenamefont {Schrieffer}}]{BCS}%
  \BibitemOpen
  \bibfield  {author} {\bibinfo {author} {\bibfnamefont {J.}~\bibnamefont
  {Bardeen}}, \bibinfo {author} {\bibfnamefont {L.~N.}\ \bibnamefont
  {Cooper}},\ and\ \bibinfo {author} {\bibfnamefont {J.~R.}\ \bibnamefont
  {Schrieffer}},\ }\bibfield  {title} {\bibinfo {title} {Theory of
  {S}uperconductivity},\ }\href {https://doi.org/10.1103/PhysRev.108.1175}
  {\bibfield  {journal} {\bibinfo  {journal} {Phys. Rev.}\ }\textbf {\bibinfo
  {volume} {108}},\ \bibinfo {pages} {1175} (\bibinfo {year}
  {1957})}\BibitemShut {NoStop}%
\bibitem [{\citenamefont {Tezuka}\ and\ \citenamefont
  {Garcia-Garcia}(2010)}]{tezuka2010}%
  \BibitemOpen
  \bibfield  {author} {\bibinfo {author} {\bibfnamefont {M.}~\bibnamefont
  {Tezuka}}\ and\ \bibinfo {author} {\bibfnamefont {A.~M.}\ \bibnamefont
  {Garcia-Garcia}},\ }\bibfield  {title} {\bibinfo {title} {Stability of the
  superfluid state in a disordered one-dimensional ultracold fermionic gas},\
  }\href@noop {} {\bibfield  {journal} {\bibinfo  {journal} {Phys. Rev. A}\
  }\textbf {\bibinfo {volume} {82}},\ \bibinfo {pages} {043613} (\bibinfo
  {year} {2010})}\BibitemShut {NoStop}%
\bibitem [{\citenamefont {Tezuka}\ and\ \citenamefont
  {Kawakami}(2013)}]{tezuka2013}%
  \BibitemOpen
  \bibfield  {author} {\bibinfo {author} {\bibfnamefont {M.}~\bibnamefont
  {Tezuka}}\ and\ \bibinfo {author} {\bibfnamefont {N.}~\bibnamefont
  {Kawakami}},\ }\bibfield  {title} {\bibinfo {title} {Reentrant topological
  transitions with {M}ajorana end states in one-dimensional superconductors by
  lattice modulation},\ }\href@noop {} {\bibfield  {journal} {\bibinfo
  {journal} {Phys. Rev. B}\ }\textbf {\bibinfo {volume} {88}},\ \bibinfo
  {pages} {155428} (\bibinfo {year} {2013})}\BibitemShut {NoStop}%
\bibitem [{\citenamefont {Cai}\ \emph {et~al.}(2013)\citenamefont {Cai},
  \citenamefont {Lang}, \citenamefont {Chen},\ and\ \citenamefont
  {Wang}}]{cai2013}%
  \BibitemOpen
  \bibfield  {author} {\bibinfo {author} {\bibfnamefont {X.}~\bibnamefont
  {Cai}}, \bibinfo {author} {\bibfnamefont {L.-J.}\ \bibnamefont {Lang}},
  \bibinfo {author} {\bibfnamefont {S.}~\bibnamefont {Chen}},\ and\ \bibinfo
  {author} {\bibfnamefont {Y.}~\bibnamefont {Wang}},\ }\bibfield  {title}
  {\bibinfo {title} {Topological {S}uperconductor to {A}nderson {L}ocalization
  {T}ransition in {O}ne-{D}imensional {I}ncommensurate {L}attices},\
  }\href@noop {} {\bibfield  {journal} {\bibinfo  {journal} {Phys. Rev. Lett.}\
  }\textbf {\bibinfo {volume} {110}},\ \bibinfo {pages} {176403} (\bibinfo
  {year} {2013})}\BibitemShut {NoStop}%
\bibitem [{\citenamefont {Fulga}\ \emph {et~al.}(2016)\citenamefont {Fulga},
  \citenamefont {Pikulin},\ and\ \citenamefont {Loring}}]{fulga2016}%
  \BibitemOpen
  \bibfield  {author} {\bibinfo {author} {\bibfnamefont {I.~C.}\ \bibnamefont
  {Fulga}}, \bibinfo {author} {\bibfnamefont {D.~I.}\ \bibnamefont {Pikulin}},\
  and\ \bibinfo {author} {\bibfnamefont {T.~A.}\ \bibnamefont {Loring}},\
  }\bibfield  {title} {\bibinfo {title} {Aperiodic {W}eak {T}opological
  {S}uperconductors},\ }\href@noop {} {\bibfield  {journal} {\bibinfo
  {journal} {Phys. Rev. Lett.}\ }\textbf {\bibinfo {volume} {116}},\ \bibinfo
  {pages} {257002} (\bibinfo {year} {2016})}\BibitemShut {NoStop}%
\bibitem [{\citenamefont {Sakai}\ \emph {et~al.}(2017)\citenamefont {Sakai},
  \citenamefont {Takemori}, \citenamefont {Koga},\ and\ \citenamefont
  {Arita}}]{Sakai2017}%
  \BibitemOpen
  \bibfield  {author} {\bibinfo {author} {\bibfnamefont {S.}~\bibnamefont
  {Sakai}}, \bibinfo {author} {\bibfnamefont {N.}~\bibnamefont {Takemori}},
  \bibinfo {author} {\bibfnamefont {A.}~\bibnamefont {Koga}},\ and\ \bibinfo
  {author} {\bibfnamefont {R.}~\bibnamefont {Arita}},\ }\bibfield  {title}
  {\bibinfo {title} {Superconductivity on a quasiperiodic lattice:
  Extended-to-localized crossover of cooper pairs},\ }\href
  {https://doi.org/10.1103/PhysRevB.95.024509} {\bibfield  {journal} {\bibinfo
  {journal} {Phys. Rev. B}\ }\textbf {\bibinfo {volume} {95}},\ \bibinfo
  {pages} {024509} (\bibinfo {year} {2017})}\BibitemShut {NoStop}%
\bibitem [{\citenamefont {Ara\'ujo}\ and\ \citenamefont
  {Andrade}(2019)}]{A-Bsuper}%
  \BibitemOpen
  \bibfield  {author} {\bibinfo {author} {\bibfnamefont {R.~N.}\ \bibnamefont
  {Ara\'ujo}}\ and\ \bibinfo {author} {\bibfnamefont {E.~C.}\ \bibnamefont
  {Andrade}},\ }\bibfield  {title} {\bibinfo {title} {Conventional
  superconductivity in quasicrystals},\ }\href
  {https://doi.org/10.1103/PhysRevB.100.014510} {\bibfield  {journal} {\bibinfo
   {journal} {Phys. Rev. B}\ }\textbf {\bibinfo {volume} {100}},\ \bibinfo
  {pages} {014510} (\bibinfo {year} {2019})}\BibitemShut {NoStop}%
\bibitem [{\citenamefont {Sakai}\ and\ \citenamefont
  {Arita}(2019)}]{sakai2019}%
  \BibitemOpen
  \bibfield  {author} {\bibinfo {author} {\bibfnamefont {S.}~\bibnamefont
  {Sakai}}\ and\ \bibinfo {author} {\bibfnamefont {R.}~\bibnamefont {Arita}},\
  }\bibfield  {title} {\bibinfo {title} {Exotic pairing state in
  quasicrystalline superconductors under a magnetic field},\ }\href@noop {}
  {\bibfield  {journal} {\bibinfo  {journal} {Phys. Rev. Res.}\ }\textbf
  {\bibinfo {volume} {1}},\ \bibinfo {pages} {022002(R)} (\bibinfo {year}
  {2019})}\BibitemShut {NoStop}%
\bibitem [{\citenamefont {Cao}\ \emph {et~al.}(2020)\citenamefont {Cao},
  \citenamefont {Zhang}, \citenamefont {Liu}, \citenamefont {Liu},
  \citenamefont {Chen},\ and\ \citenamefont {Yang}}]{cao2020}%
  \BibitemOpen
  \bibfield  {author} {\bibinfo {author} {\bibfnamefont {Y.}~\bibnamefont
  {Cao}}, \bibinfo {author} {\bibfnamefont {Y.}~\bibnamefont {Zhang}}, \bibinfo
  {author} {\bibfnamefont {Y.-B.}\ \bibnamefont {Liu}}, \bibinfo {author}
  {\bibfnamefont {C.-C.}\ \bibnamefont {Liu}}, \bibinfo {author} {\bibfnamefont
  {W.-Q.}\ \bibnamefont {Chen}},\ and\ \bibinfo {author} {\bibfnamefont
  {F.}~\bibnamefont {Yang}},\ }\bibfield  {title} {\bibinfo {title}
  {Kohn-{L}uttinger {M}echanism {D}riven {E}xotic {T}opological
  {S}uperconductivity on the {P}enrose {L}attice},\ }\href@noop {} {\bibfield
  {journal} {\bibinfo  {journal} {Phys. Rev. Lett.}\ }\textbf {\bibinfo
  {volume} {125}},\ \bibinfo {pages} {017002} (\bibinfo {year}
  {2020})}\BibitemShut {NoStop}%
\bibitem [{\citenamefont {Takemori}\ \emph {et~al.}(2020)\citenamefont
  {Takemori}, \citenamefont {Arita},\ and\ \citenamefont {Sakai}}]{Takemorisc}%
  \BibitemOpen
  \bibfield  {author} {\bibinfo {author} {\bibfnamefont {N.}~\bibnamefont
  {Takemori}}, \bibinfo {author} {\bibfnamefont {R.}~\bibnamefont {Arita}},\
  and\ \bibinfo {author} {\bibfnamefont {S.}~\bibnamefont {Sakai}},\ }\bibfield
   {title} {\bibinfo {title} {Physical properties of weak-coupling
  quasiperiodic superconductors},\ }\href
  {https://doi.org/10.1103/PhysRevB.102.115108} {\bibfield  {journal} {\bibinfo
   {journal} {Phys. Rev. B}\ }\textbf {\bibinfo {volume} {102}},\ \bibinfo
  {pages} {115108} (\bibinfo {year} {2020})}\BibitemShut {NoStop}%
\bibitem [{\citenamefont {Ghadimi}\ \emph {et~al.}(2021)\citenamefont
  {Ghadimi}, \citenamefont {Sugimoto}, \citenamefont {Tanaka},\ and\
  \citenamefont {Tohyama}}]{ghadimi2021}%
  \BibitemOpen
  \bibfield  {author} {\bibinfo {author} {\bibfnamefont {R.}~\bibnamefont
  {Ghadimi}}, \bibinfo {author} {\bibfnamefont {T.}~\bibnamefont {Sugimoto}},
  \bibinfo {author} {\bibfnamefont {K.}~\bibnamefont {Tanaka}},\ and\ \bibinfo
  {author} {\bibfnamefont {T.}~\bibnamefont {Tohyama}},\ }\bibfield  {title}
  {\bibinfo {title} {Topological superconductivity in quasicrystals},\
  }\href@noop {} {\bibfield  {journal} {\bibinfo  {journal} {Phys. Rev. B}\
  }\textbf {\bibinfo {volume} {104}},\ \bibinfo {pages} {144511} (\bibinfo
  {year} {2021})}\BibitemShut {NoStop}%
\bibitem [{\citenamefont {Liu}\ \emph {et~al.}(2022)\citenamefont {Liu},
  \citenamefont {Hao}, \citenamefont {Zhang}, \citenamefont {Cao},
  \citenamefont {Chen},\ and\ \citenamefont {Yang}}]{liu2022cooper}%
  \BibitemOpen
  \bibfield  {author} {\bibinfo {author} {\bibfnamefont {Y.-B.}\ \bibnamefont
  {Liu}}, \bibinfo {author} {\bibfnamefont {J.-J.}\ \bibnamefont {Hao}},
  \bibinfo {author} {\bibfnamefont {Y.}~\bibnamefont {Zhang}}, \bibinfo
  {author} {\bibfnamefont {Y.}~\bibnamefont {Cao}}, \bibinfo {author}
  {\bibfnamefont {W.-Q.}\ \bibnamefont {Chen}},\ and\ \bibinfo {author}
  {\bibfnamefont {F.}~\bibnamefont {Yang}},\ }\bibfield  {title} {\bibinfo
  {title} {Cooper instability and superconductivity of the {P}enrose lattice},\
  }\href@noop {} {\bibfield  {journal} {\bibinfo  {journal} {Sci. China-Phys.
  Mech. Astron.}\ }\textbf {\bibinfo {volume} {65}},\ \bibinfo {pages} {1}
  (\bibinfo {year} {2022})}\BibitemShut {NoStop}%
\bibitem [{\citenamefont {Uri}\ \emph {et~al.}(2023)\citenamefont {Uri},
  \citenamefont {de~la Barrera}, \citenamefont {Randeria}, \citenamefont
  {Rodan-Legrain}, \citenamefont {Devakul}, \citenamefont {Crowley},
  \citenamefont {Paul}, \citenamefont {Watanabe}, \citenamefont {Taniguchi},
  \citenamefont {Lifshitz}, \citenamefont {Fu}, \citenamefont {Ashoori},\ and\
  \citenamefont {Jarillo-Herrero}}]{Uri23}%
  \BibitemOpen
  \bibfield  {author} {\bibinfo {author} {\bibfnamefont {A.}~\bibnamefont
  {Uri}}, \bibinfo {author} {\bibfnamefont {S.~C.}\ \bibnamefont {de~la
  Barrera}}, \bibinfo {author} {\bibfnamefont {M.~T.}\ \bibnamefont
  {Randeria}}, \bibinfo {author} {\bibfnamefont {D.}~\bibnamefont
  {Rodan-Legrain}}, \bibinfo {author} {\bibfnamefont {T.}~\bibnamefont
  {Devakul}}, \bibinfo {author} {\bibfnamefont {P.~J.~D.}\ \bibnamefont
  {Crowley}}, \bibinfo {author} {\bibfnamefont {N.}~\bibnamefont {Paul}},
  \bibinfo {author} {\bibfnamefont {K.}~\bibnamefont {Watanabe}}, \bibinfo
  {author} {\bibfnamefont {T.}~\bibnamefont {Taniguchi}}, \bibinfo {author}
  {\bibfnamefont {R.}~\bibnamefont {Lifshitz}}, \bibinfo {author}
  {\bibfnamefont {L.}~\bibnamefont {Fu}}, \bibinfo {author} {\bibfnamefont
  {R.~C.}\ \bibnamefont {Ashoori}},\ and\ \bibinfo {author} {\bibfnamefont
  {P.}~\bibnamefont {Jarillo-Herrero}},\ }\bibfield  {title} {\bibinfo {title}
  {Superconductivity and strong interactions in a tunable moiré
  quasicrystal},\ }\href {https://doi.org/10.1038/s41586-023-06294-z}
  {\bibfield  {journal} {\bibinfo  {journal} {Nature}\ }\textbf {\bibinfo
  {volume} {620}},\ \bibinfo {pages} {762} (\bibinfo {year}
  {2023})}\BibitemShut {NoStop}%
\bibitem [{\citenamefont {Misko}\ \emph {et~al.}(2005)\citenamefont {Misko},
  \citenamefont {Savel'ev},\ and\ \citenamefont {Nori}}]{Misko05}%
  \BibitemOpen
  \bibfield  {author} {\bibinfo {author} {\bibfnamefont {V.}~\bibnamefont
  {Misko}}, \bibinfo {author} {\bibfnamefont {S.}~\bibnamefont {Savel'ev}},\
  and\ \bibinfo {author} {\bibfnamefont {F.}~\bibnamefont {Nori}},\ }\bibfield
  {title} {\bibinfo {title} {Critical currents in quasiperiodic pinning arrays:
  Chains and penrose lattices},\ }\href
  {https://doi.org/10.1103/PhysRevLett.95.177007} {\bibfield  {journal}
  {\bibinfo  {journal} {Phys. Rev. Lett.}\ }\textbf {\bibinfo {volume} {95}},\
  \bibinfo {pages} {177007} (\bibinfo {year} {2005})}\BibitemShut {NoStop}%
\bibitem [{\citenamefont {Misko}\ \emph {et~al.}(2006)\citenamefont {Misko},
  \citenamefont {Savel'ev},\ and\ \citenamefont {Nori}}]{Misko06}%
  \BibitemOpen
  \bibfield  {author} {\bibinfo {author} {\bibfnamefont {V.~R.}\ \bibnamefont
  {Misko}}, \bibinfo {author} {\bibfnamefont {S.}~\bibnamefont {Savel'ev}},\
  and\ \bibinfo {author} {\bibfnamefont {F.}~\bibnamefont {Nori}},\ }\bibfield
  {title} {\bibinfo {title} {Critical currents in superconductors with
  quasiperiodic pinning arrays: One-dimensional chains and two-dimensional
  penrose lattices},\ }\href {https://doi.org/10.1103/PhysRevB.74.024522}
  {\bibfield  {journal} {\bibinfo  {journal} {Phys. Rev. B}\ }\textbf {\bibinfo
  {volume} {74}},\ \bibinfo {pages} {024522} (\bibinfo {year}
  {2006})}\BibitemShut {NoStop}%
\bibitem [{\citenamefont {Kemmler}\ \emph {et~al.}(2006)\citenamefont
  {Kemmler}, \citenamefont {G\"urlich}, \citenamefont {Sterck}, \citenamefont
  {P\"ohler}, \citenamefont {Neuhaus}, \citenamefont {Siegel}, \citenamefont
  {Kleiner},\ and\ \citenamefont {Koelle}}]{Kemmler06}%
  \BibitemOpen
  \bibfield  {author} {\bibinfo {author} {\bibfnamefont {M.}~\bibnamefont
  {Kemmler}}, \bibinfo {author} {\bibfnamefont {C.}~\bibnamefont {G\"urlich}},
  \bibinfo {author} {\bibfnamefont {A.}~\bibnamefont {Sterck}}, \bibinfo
  {author} {\bibfnamefont {H.}~\bibnamefont {P\"ohler}}, \bibinfo {author}
  {\bibfnamefont {M.}~\bibnamefont {Neuhaus}}, \bibinfo {author} {\bibfnamefont
  {M.}~\bibnamefont {Siegel}}, \bibinfo {author} {\bibfnamefont
  {R.}~\bibnamefont {Kleiner}},\ and\ \bibinfo {author} {\bibfnamefont
  {D.}~\bibnamefont {Koelle}},\ }\bibfield  {title} {\bibinfo {title}
  {Commensurability effects in superconducting nb films with quasiperiodic
  pinning arrays},\ }\href {https://doi.org/10.1103/PhysRevLett.97.147003}
  {\bibfield  {journal} {\bibinfo  {journal} {Phys. Rev. Lett.}\ }\textbf
  {\bibinfo {volume} {97}},\ \bibinfo {pages} {147003} (\bibinfo {year}
  {2006})}\BibitemShut {NoStop}%
\bibitem [{\citenamefont {Silhanek}\ \emph {et~al.}(2006)\citenamefont
  {Silhanek}, \citenamefont {Gillijns}, \citenamefont {Moshchalkov},
  \citenamefont {Zhu}, \citenamefont {Moonens},\ and\ \citenamefont
  {Leunissen}}]{Silhanek06}%
  \BibitemOpen
  \bibfield  {author} {\bibinfo {author} {\bibfnamefont {A.}~\bibnamefont
  {Silhanek}}, \bibinfo {author} {\bibfnamefont {W.}~\bibnamefont {Gillijns}},
  \bibinfo {author} {\bibfnamefont {V.}~\bibnamefont {Moshchalkov}}, \bibinfo
  {author} {\bibfnamefont {B.}~\bibnamefont {Zhu}}, \bibinfo {author}
  {\bibfnamefont {J.}~\bibnamefont {Moonens}},\ and\ \bibinfo {author}
  {\bibfnamefont {L.}~\bibnamefont {Leunissen}},\ }\bibfield  {title} {\bibinfo
  {title} {Enhanced pinning and proliferation of matching effects in a
  superconducting film with a penrose array of magnetic dots},\ }\href@noop {}
  {\bibfield  {journal} {\bibinfo  {journal} {Applied physics letters}\
  }\textbf {\bibinfo {volume} {89}},\ \bibinfo {pages} {152507} (\bibinfo
  {year} {2006})}\BibitemShut {NoStop}%
\bibitem [{\citenamefont {Misko}\ \emph {et~al.}(2010)\citenamefont {Misko},
  \citenamefont {Bothner}, \citenamefont {Kemmler}, \citenamefont {Kleiner},
  \citenamefont {Koelle}, \citenamefont {Peeters},\ and\ \citenamefont
  {Nori}}]{Misko10}%
  \BibitemOpen
  \bibfield  {author} {\bibinfo {author} {\bibfnamefont {V.~R.}\ \bibnamefont
  {Misko}}, \bibinfo {author} {\bibfnamefont {D.}~\bibnamefont {Bothner}},
  \bibinfo {author} {\bibfnamefont {M.}~\bibnamefont {Kemmler}}, \bibinfo
  {author} {\bibfnamefont {R.}~\bibnamefont {Kleiner}}, \bibinfo {author}
  {\bibfnamefont {D.}~\bibnamefont {Koelle}}, \bibinfo {author} {\bibfnamefont
  {F.~M.}\ \bibnamefont {Peeters}},\ and\ \bibinfo {author} {\bibfnamefont
  {F.}~\bibnamefont {Nori}},\ }\bibfield  {title} {\bibinfo {title} {Enhancing
  the critical current in quasiperiodic pinning arrays below and above the
  matching magnetic flux},\ }\href {https://doi.org/10.1103/PhysRevB.82.184512}
  {\bibfield  {journal} {\bibinfo  {journal} {Phys. Rev. B}\ }\textbf {\bibinfo
  {volume} {82}},\ \bibinfo {pages} {184512} (\bibinfo {year}
  {2010})}\BibitemShut {NoStop}%
\bibitem [{\citenamefont {Gordon}\ \emph {et~al.}(1986)\citenamefont {Gordon},
  \citenamefont {Goldman}, \citenamefont {Maps}, \citenamefont {Costello},
  \citenamefont {Tiberio},\ and\ \citenamefont {Whitehead}}]{Gordon86}%
  \BibitemOpen
  \bibfield  {author} {\bibinfo {author} {\bibfnamefont {J.~M.}\ \bibnamefont
  {Gordon}}, \bibinfo {author} {\bibfnamefont {A.~M.}\ \bibnamefont {Goldman}},
  \bibinfo {author} {\bibfnamefont {J.}~\bibnamefont {Maps}}, \bibinfo {author}
  {\bibfnamefont {D.}~\bibnamefont {Costello}}, \bibinfo {author}
  {\bibfnamefont {R.}~\bibnamefont {Tiberio}},\ and\ \bibinfo {author}
  {\bibfnamefont {B.}~\bibnamefont {Whitehead}},\ }\bibfield  {title} {\bibinfo
  {title} {Superconducting-normal phase boundary of a fractal network in a
  magnetic field},\ }\href {https://doi.org/10.1103/PhysRevLett.56.2280}
  {\bibfield  {journal} {\bibinfo  {journal} {Phys. Rev. Lett.}\ }\textbf
  {\bibinfo {volume} {56}},\ \bibinfo {pages} {2280} (\bibinfo {year}
  {1986})}\BibitemShut {NoStop}%
\bibitem [{\citenamefont {Behrooz}\ \emph {et~al.}(1986)\citenamefont
  {Behrooz}, \citenamefont {Burns}, \citenamefont {Deckman}, \citenamefont
  {Levine}, \citenamefont {Whitehead},\ and\ \citenamefont
  {Chaikin}}]{Behrooz86}%
  \BibitemOpen
  \bibfield  {author} {\bibinfo {author} {\bibfnamefont {A.}~\bibnamefont
  {Behrooz}}, \bibinfo {author} {\bibfnamefont {M.~J.}\ \bibnamefont {Burns}},
  \bibinfo {author} {\bibfnamefont {H.}~\bibnamefont {Deckman}}, \bibinfo
  {author} {\bibfnamefont {D.}~\bibnamefont {Levine}}, \bibinfo {author}
  {\bibfnamefont {B.}~\bibnamefont {Whitehead}},\ and\ \bibinfo {author}
  {\bibfnamefont {P.~M.}\ \bibnamefont {Chaikin}},\ }\bibfield  {title}
  {\bibinfo {title} {Flux quantization on quasicrystalline networks},\ }\href
  {https://doi.org/10.1103/PhysRevLett.57.368} {\bibfield  {journal} {\bibinfo
  {journal} {Phys. Rev. Lett.}\ }\textbf {\bibinfo {volume} {57}},\ \bibinfo
  {pages} {368} (\bibinfo {year} {1986})}\BibitemShut {NoStop}%
\bibitem [{\citenamefont {Springer}\ and\ \citenamefont
  {Van~Harlingen}(1987)}]{Springer87}%
  \BibitemOpen
  \bibfield  {author} {\bibinfo {author} {\bibfnamefont {K.~N.}\ \bibnamefont
  {Springer}}\ and\ \bibinfo {author} {\bibfnamefont {D.~J.}\ \bibnamefont
  {Van~Harlingen}},\ }\bibfield  {title} {\bibinfo {title} {Resistive
  transition and magnetic field response of a penrose-tile array of weakly
  coupled superconductor islands},\ }\href
  {https://doi.org/10.1103/PhysRevB.36.7273} {\bibfield  {journal} {\bibinfo
  {journal} {Phys. Rev. B}\ }\textbf {\bibinfo {volume} {36}},\ \bibinfo
  {pages} {7273} (\bibinfo {year} {1987})}\BibitemShut {NoStop}%
\bibitem [{\citenamefont {Nori}\ \emph {et~al.}(1987)\citenamefont {Nori},
  \citenamefont {Niu}, \citenamefont {Fradkin},\ and\ \citenamefont
  {Chang}}]{Nori87}%
  \BibitemOpen
  \bibfield  {author} {\bibinfo {author} {\bibfnamefont {F.}~\bibnamefont
  {Nori}}, \bibinfo {author} {\bibfnamefont {Q.}~\bibnamefont {Niu}}, \bibinfo
  {author} {\bibfnamefont {E.}~\bibnamefont {Fradkin}},\ and\ \bibinfo {author}
  {\bibfnamefont {S.-J.}\ \bibnamefont {Chang}},\ }\bibfield  {title} {\bibinfo
  {title} {Superconducting-normal phase boundary of quasicrystalline arrays in
  a magnetic field},\ }\href {https://doi.org/10.1103/PhysRevB.36.8338}
  {\bibfield  {journal} {\bibinfo  {journal} {Phys. Rev. B}\ }\textbf {\bibinfo
  {volume} {36}},\ \bibinfo {pages} {8338} (\bibinfo {year}
  {1987})}\BibitemShut {NoStop}%
\bibitem [{\citenamefont {Nori}\ and\ \citenamefont {Niu}(1988)}]{Nori88}%
  \BibitemOpen
  \bibfield  {author} {\bibinfo {author} {\bibfnamefont {F.}~\bibnamefont
  {Nori}}\ and\ \bibinfo {author} {\bibfnamefont {Q.}~\bibnamefont {Niu}},\
  }\bibfield  {title} {\bibinfo {title} {Tc(h) for quasicrystalline
  micronetworks: Analytical and numerical results},\ }\href
  {https://doi.org/https://doi.org/10.1016/0921-4526(88)90075-0} {\bibfield
  {journal} {\bibinfo  {journal} {Physica B: Condensed Matter}\ }\textbf
  {\bibinfo {volume} {152}},\ \bibinfo {pages} {105 } (\bibinfo {year}
  {1988})}\BibitemShut {NoStop}%
\bibitem [{\citenamefont {Niu}\ and\ \citenamefont {Nori}(1989)}]{Niu89}%
  \BibitemOpen
  \bibfield  {author} {\bibinfo {author} {\bibfnamefont {Q.}~\bibnamefont
  {Niu}}\ and\ \bibinfo {author} {\bibfnamefont {F.}~\bibnamefont {Nori}},\
  }\bibfield  {title} {\bibinfo {title} {Theory of superconducting wire
  networks and josephson-junction arrays in magnetic fields},\ }\href
  {https://doi.org/10.1103/PhysRevB.39.2134} {\bibfield  {journal} {\bibinfo
  {journal} {Phys. Rev. B}\ }\textbf {\bibinfo {volume} {39}},\ \bibinfo
  {pages} {2134} (\bibinfo {year} {1989})}\BibitemShut {NoStop}%
\bibitem [{\citenamefont {Schrieffer}(2018)}]{schrieffer}%
  \BibitemOpen
  \bibfield  {author} {\bibinfo {author} {\bibfnamefont {J.~R.}\ \bibnamefont
  {Schrieffer}},\ }\href@noop {} {\emph {\bibinfo {title} {Theory of
  superconductivity}}}\ (\bibinfo  {publisher} {CRC press},\ \bibinfo {year}
  {2018})\ Chap.~\bibinfo {chapter} {8}\BibitemShut {NoStop}%
\bibitem [{\citenamefont {Beenker}(1982)}]{beenker}%
  \BibitemOpen
  \bibfield  {author} {\bibinfo {author} {\bibfnamefont {F.~P.~M.}\
  \bibnamefont {Beenker}},\ }\href@noop {} {\emph {\bibinfo {title} {Algebraic
  theory of non-periodic tilings of the plane by two simple building blocks: a
  square and a rhombus}}},\ \bibinfo {type} {Tech. Rep.}\ \bibinfo {number}
  {82-WSK04}\ (\bibinfo  {institution} {Eindhoven University of Technology},\
  \bibinfo {year} {1982})\BibitemShut {NoStop}%
\bibitem [{\citenamefont {Socolar}(1989)}]{AB}%
  \BibitemOpen
  \bibfield  {author} {\bibinfo {author} {\bibfnamefont {J.~E.~S.}\
  \bibnamefont {Socolar}},\ }\bibfield  {title} {\bibinfo {title} {Simple
  octagonal and dodecagonal quasicrystals},\ }\href
  {https://doi.org/10.1103/PhysRevB.39.10519} {\bibfield  {journal} {\bibinfo
  {journal} {Phys. Rev. B}\ }\textbf {\bibinfo {volume} {39}},\ \bibinfo
  {pages} {10519} (\bibinfo {year} {1989})}\BibitemShut {NoStop}%
\bibitem [{\citenamefont {Ammann}\ \emph {et~al.}(1992)\citenamefont {Ammann},
  \citenamefont {Gr{\"u}nbaum},\ and\ \citenamefont {Shephard}}]{ammann}%
  \BibitemOpen
  \bibfield  {author} {\bibinfo {author} {\bibfnamefont {R.}~\bibnamefont
  {Ammann}}, \bibinfo {author} {\bibfnamefont {B.}~\bibnamefont
  {Gr{\"u}nbaum}},\ and\ \bibinfo {author} {\bibfnamefont {G.~C.}\ \bibnamefont
  {Shephard}},\ }\bibfield  {title} {\bibinfo {title} {Aperiodic tiles},\
  }\href@noop {} {\bibfield  {journal} {\bibinfo  {journal} {Discrete Comput.
  Geom.}\ }\textbf {\bibinfo {volume} {8}},\ \bibinfo {pages} {1} (\bibinfo
  {year} {1992})}\BibitemShut {NoStop}%
\bibitem [{\citenamefont {Fukushima}\ \emph {et~al.}(2023)\citenamefont
  {Fukushima}, \citenamefont {Takemori}, \citenamefont {Sakai}, \citenamefont
  {Ichioka},\ and\ \citenamefont {Jagannathan}}]{TF}%
  \BibitemOpen
  \bibfield  {author} {\bibinfo {author} {\bibfnamefont {T.}~\bibnamefont
  {Fukushima}}, \bibinfo {author} {\bibfnamefont {N.}~\bibnamefont {Takemori}},
  \bibinfo {author} {\bibfnamefont {S.}~\bibnamefont {Sakai}}, \bibinfo
  {author} {\bibfnamefont {M.}~\bibnamefont {Ichioka}},\ and\ \bibinfo {author}
  {\bibfnamefont {A.}~\bibnamefont {Jagannathan}},\ }\bibfield  {title}
  {\bibinfo {title} {Supercurrent {D}istribution on {A}mmann-{B}eenker
  {S}tructure},\ }\href {https://doi.org/10.1088/1742-6596/2461/1/012014}
  {\bibfield  {journal} {\bibinfo  {journal} {J. Phys.: Conf. Ser.}\ }\textbf
  {\bibinfo {volume} {2461}},\ \bibinfo {pages} {012014} (\bibinfo {year}
  {2023})}\BibitemShut {NoStop}%
\bibitem [{\citenamefont {Duneau}\ \emph {et~al.}(1989)\citenamefont {Duneau},
  \citenamefont {Mosseri},\ and\ \citenamefont {Oguey}}]{duneau}%
  \BibitemOpen
  \bibfield  {author} {\bibinfo {author} {\bibfnamefont {M.}~\bibnamefont
  {Duneau}}, \bibinfo {author} {\bibfnamefont {R.}~\bibnamefont {Mosseri}},\
  and\ \bibinfo {author} {\bibfnamefont {C.}~\bibnamefont {Oguey}},\ }\bibfield
   {title} {\bibinfo {title} {Approximants of quasiperiodic structures
  generated by the inflation mapping},\ }\href@noop {} {\bibfield  {journal}
  {\bibinfo  {journal} {J. Phys. A}\ }\textbf {\bibinfo {volume} {22}},\
  \bibinfo {pages} {4549} (\bibinfo {year} {1989})}\BibitemShut {NoStop}%
\bibitem [{\citenamefont {Jagannathan}(2005)}]{Jag2005}%
  \BibitemOpen
  \bibfield  {author} {\bibinfo {author} {\bibfnamefont {A.}~\bibnamefont
  {Jagannathan}},\ }\bibfield  {title} {\bibinfo {title} {Ground state of a
  two-dimensional quasiperiodic quantum antiferromagnet},\ }\href
  {https://doi.org/10.1103/PhysRevB.71.115101} {\bibfield  {journal} {\bibinfo
  {journal} {Phys. Rev. B}\ }\textbf {\bibinfo {volume} {71}},\ \bibinfo
  {pages} {115101} (\bibinfo {year} {2005})}\BibitemShut {NoStop}%
\bibitem [{\citenamefont {Jagannathan}(2004)}]{Jag2004}%
  \BibitemOpen
  \bibfield  {author} {\bibinfo {author} {\bibfnamefont {A.}~\bibnamefont
  {Jagannathan}},\ }\bibfield  {title} {\bibinfo {title} {Quantum {S}pins and
  {Q}uasiperiodicity: {A} {R}eal {S}pace {R}enormalization {G}roup
  {A}pproach},\ }\href {https://doi.org/10.1103/PhysRevLett.92.047202}
  {\bibfield  {journal} {\bibinfo  {journal} {Phys. Rev. Lett.}\ }\textbf
  {\bibinfo {volume} {92}},\ \bibinfo {pages} {047202} (\bibinfo {year}
  {2004})}\BibitemShut {NoStop}%
\bibitem [{\citenamefont {Esslinger}(2010)}]{Esslinger2010}%
  \BibitemOpen
  \bibfield  {author} {\bibinfo {author} {\bibfnamefont {T.}~\bibnamefont
  {Esslinger}},\ }\bibfield  {title} {\bibinfo {title} {Fermi-{H}ubbard
  {P}hysics with {A}toms in an {O}ptical {L}attice},\ }\href
  {https://doi.org/10.1146/annurev-conmatphys-070909-104059} {\bibfield
  {journal} {\bibinfo  {journal} {Annu. Rev. Condens. Matter Phys.}\ }\textbf
  {\bibinfo {volume} {1}},\ \bibinfo {pages} {129} (\bibinfo {year}
  {2010})}\BibitemShut {NoStop}%
\bibitem [{\citenamefont {Peierls}(1997)}]{peierls}%
  \BibitemOpen
  \bibfield  {author} {\bibinfo {author} {\bibfnamefont {R.}~\bibnamefont
  {Peierls}},\ }\bibfield  {title} {\bibinfo {title} {On the {T}heory of the
  {D}iamagnetism of {C}onduction {E}lectrons},\ }in\ \href@noop {} {\emph
  {\bibinfo {booktitle} {Selected Scientific Papers of Sir Rudolf Peierls:
  (With Commentary)}}}\ (\bibinfo  {publisher} {World Scientific},\ \bibinfo
  {year} {1997})\ pp.\ \bibinfo {pages} {97--120}\BibitemShut {NoStop}%
\bibitem [{\citenamefont {Sakai}\ and\ \citenamefont {Koga}(2021)}]{sakai2021}%
  \BibitemOpen
  \bibfield  {author} {\bibinfo {author} {\bibfnamefont {S.}~\bibnamefont
  {Sakai}}\ and\ \bibinfo {author} {\bibfnamefont {A.}~\bibnamefont {Koga}},\
  }\bibfield  {title} {\bibinfo {title} {Effect of {E}lectron-{E}lectron
  {I}nteractions on {M}etallic {S}tate in {Q}uasicrystals},\ }\href@noop {}
  {\bibfield  {journal} {\bibinfo  {journal} {Mater. Trans.}\ }\textbf
  {\bibinfo {volume} {62}},\ \bibinfo {pages} {380} (\bibinfo {year}
  {2021})}\BibitemShut {NoStop}%
\bibitem [{\citenamefont {Koga}(2020)}]{KogaAB}%
  \BibitemOpen
  \bibfield  {author} {\bibinfo {author} {\bibfnamefont {A.}~\bibnamefont
  {Koga}},\ }\bibfield  {title} {\bibinfo {title} {Superlattice structure in
  the antiferromagnetically ordered state in the {H}ubbard model on the
  {A}mmann-{B}eenker tiling},\ }\href
  {https://doi.org/10.1103/PhysRevB.102.115125} {\bibfield  {journal} {\bibinfo
   {journal} {Phys. Rev. B}\ }\textbf {\bibinfo {volume} {102}},\ \bibinfo
  {pages} {115125} (\bibinfo {year} {2020})}\BibitemShut {NoStop}%
\bibitem [{\citenamefont {Dobrosavljevic}\ \emph {et~al.}(2012)\citenamefont
  {Dobrosavljevic}, \citenamefont {Trivedi},\ and\ \citenamefont
  {Valles~Jr}}]{dobrosavljevic}%
  \BibitemOpen
  \bibfield  {author} {\bibinfo {author} {\bibfnamefont {V.}~\bibnamefont
  {Dobrosavljevic}}, \bibinfo {author} {\bibfnamefont {N.}~\bibnamefont
  {Trivedi}},\ and\ \bibinfo {author} {\bibfnamefont {J.~M.}\ \bibnamefont
  {Valles~Jr}},\ }\href@noop {} {\emph {\bibinfo {title} {Conductor-{I}nsulator
  {Q}uantum {P}hase {T}ransitions}}}\ (\bibinfo  {publisher} {Oxford University
  Press},\ \bibinfo {year} {2012})\BibitemShut {NoStop}%
\bibitem [{\citenamefont {Kita}(2015)}]{kita}%
  \BibitemOpen
  \bibfield  {author} {\bibinfo {author} {\bibfnamefont {T.}~\bibnamefont
  {Kita}},\ }\href@noop {} {\emph {\bibinfo {title} {Statistical Mechanics of
  Superconductivity}}}\ (\bibinfo  {publisher} {Springer},\ \bibinfo {year}
  {2015})\BibitemShut {NoStop}%
\bibitem [{\citenamefont {De~Gennes}(1999)}]{de2018superconductivity}%
  \BibitemOpen
  \bibfield  {author} {\bibinfo {author} {\bibfnamefont {P.~G.}\ \bibnamefont
  {De~Gennes}},\ }\href {https://cds.cern.ch/record/566105} {\emph {\bibinfo
  {title} {{Superconductivity of Metals and Alloys}}}},\ Advanced book
  classics\ (\bibinfo  {publisher} {Perseus},\ \bibinfo {address} {Cambridge,
  MA},\ \bibinfo {year} {1999})\BibitemShut {NoStop}%
\bibitem [{\citenamefont {Nagai}(2020)}]{Nagai2020}%
  \BibitemOpen
  \bibfield  {author} {\bibinfo {author} {\bibfnamefont {Y.}~\bibnamefont
  {Nagai}},\ }\bibfield  {title} {\bibinfo {title} {N-independent {L}ocalized
  {K}rylov–{B}ogoliubov-de {G}ennes {M}ethod: {U}ltra-fast {N}umerical
  {A}pproach to {L}arge-scale {I}nhomogeneous {S}uperconductors},\ }\href
  {https://doi.org/10.7566/JPSJ.89.074703} {\bibfield  {journal} {\bibinfo
  {journal} {J. Phys. Soc. Japan.}\ }\textbf {\bibinfo {volume} {89}},\
  \bibinfo {pages} {074703} (\bibinfo {year} {2020})}\BibitemShut {NoStop}%
\bibitem [{\citenamefont {Nagai}(2022)}]{PhysRevB.106.064506}%
  \BibitemOpen
  \bibfield  {author} {\bibinfo {author} {\bibfnamefont {Y.}~\bibnamefont
  {Nagai}},\ }\bibfield  {title} {\bibinfo {title} {Intrinsic vortex pinning in
  superconducting quasicrystals},\ }\href
  {https://doi.org/10.1103/PhysRevB.106.064506} {\bibfield  {journal} {\bibinfo
   {journal} {Phys. Rev. B}\ }\textbf {\bibinfo {volume} {106}},\ \bibinfo
  {pages} {064506} (\bibinfo {year} {2022})}\BibitemShut {NoStop}%
\bibitem [{\citenamefont {Nabeta}\ \emph {et~al.}(2017)\citenamefont {Nabeta},
  \citenamefont {Tanaka}, \citenamefont {Onari},\ and\ \citenamefont
  {Ichioka}}]{PhysRevB.96.094522}%
  \BibitemOpen
  \bibfield  {author} {\bibinfo {author} {\bibfnamefont {M.}~\bibnamefont
  {Nabeta}}, \bibinfo {author} {\bibfnamefont {K.~K.}\ \bibnamefont {Tanaka}},
  \bibinfo {author} {\bibfnamefont {S.}~\bibnamefont {Onari}},\ and\ \bibinfo
  {author} {\bibfnamefont {M.}~\bibnamefont {Ichioka}},\ }\bibfield  {title}
  {\bibinfo {title} {Pair breaking of multigap superconductivity under parallel
  magnetic fields in the electric-field-induced surface metallic state},\
  }\href {https://doi.org/10.1103/PhysRevB.96.094522} {\bibfield  {journal}
  {\bibinfo  {journal} {Phys. Rev. B}\ }\textbf {\bibinfo {volume} {96}},\
  \bibinfo {pages} {094522} (\bibinfo {year} {2017})}\BibitemShut {NoStop}%
\bibitem [{\citenamefont {Takigawa}\ \emph {et~al.}(2001)\citenamefont
  {Takigawa}, \citenamefont {Ichioka}, \citenamefont {Machida},\ and\
  \citenamefont {Sigrist}}]{PhysRevB.65.014508}%
  \BibitemOpen
  \bibfield  {author} {\bibinfo {author} {\bibfnamefont {M.}~\bibnamefont
  {Takigawa}}, \bibinfo {author} {\bibfnamefont {M.}~\bibnamefont {Ichioka}},
  \bibinfo {author} {\bibfnamefont {K.}~\bibnamefont {Machida}},\ and\ \bibinfo
  {author} {\bibfnamefont {M.}~\bibnamefont {Sigrist}},\ }\bibfield  {title}
  {\bibinfo {title} {Vortex structure in chiral $p$-wave superconductors},\
  }\href {https://doi.org/10.1103/PhysRevB.65.014508} {\bibfield  {journal}
  {\bibinfo  {journal} {Phys. Rev. B}\ }\textbf {\bibinfo {volume} {65}},\
  \bibinfo {pages} {014508} (\bibinfo {year} {2001})}\BibitemShut {NoStop}%
\bibitem [{\citenamefont {Jagannathan}(1994)}]{aj1994}%
  \BibitemOpen
  \bibfield  {author} {\bibinfo {author} {\bibfnamefont {A.}~\bibnamefont
  {Jagannathan}},\ }\bibfield  {title} {\bibinfo {title} {Density of states and
  magnetic susceptibilities on the octagonal tiling},\ }\href@noop {}
  {\bibfield  {journal} {\bibinfo  {journal} {J. Phys. I (France)}\ }\textbf
  {\bibinfo {volume} {4}},\ \bibinfo {pages} {133} (\bibinfo {year}
  {1994})}\BibitemShut {NoStop}%
\bibitem [{\citenamefont {Jagannathan}(2023)}]{aj2023}%
  \BibitemOpen
  \bibfield  {author} {\bibinfo {author} {\bibfnamefont {A.}~\bibnamefont
  {Jagannathan}},\ }\bibfield  {title} {\bibinfo {title} {Closing of gaps and
  gap labeling and passage from molecular states to critical states in a 2d
  quasicrystal},\ }\href@noop {} {\bibfield  {journal} {\bibinfo  {journal}
  {arXiv preprint arXiv:2304.04409}\ } (\bibinfo {year} {2023})}\BibitemShut
  {NoStop}%
\bibitem [{\citenamefont {Fulde}\ and\ \citenamefont {Ferrell}(1964)}]{FFLO1}%
  \BibitemOpen
  \bibfield  {author} {\bibinfo {author} {\bibfnamefont {P.}~\bibnamefont
  {Fulde}}\ and\ \bibinfo {author} {\bibfnamefont {R.~A.}\ \bibnamefont
  {Ferrell}},\ }\bibfield  {title} {\bibinfo {title} {Superconductivity in a
  {S}trong {S}pin-{E}xchange {F}ield},\ }\href
  {https://doi.org/10.1103/PhysRev.135.A550} {\bibfield  {journal} {\bibinfo
  {journal} {Phys. Rev.}\ }\textbf {\bibinfo {volume} {135}},\ \bibinfo {pages}
  {A550} (\bibinfo {year} {1964})}\BibitemShut {NoStop}%
\bibitem [{\citenamefont {Larkin}\ and\ \citenamefont
  {Ovchinnikov}(1965)}]{FFLO2}%
  \BibitemOpen
  \bibfield  {author} {\bibinfo {author} {\bibfnamefont {A.}~\bibnamefont
  {Larkin}}\ and\ \bibinfo {author} {\bibfnamefont {Y.~N.}\ \bibnamefont
  {Ovchinnikov}},\ }\bibfield  {title} {\bibinfo {title} {Nonuniform state of
  superconductors},\ }\href@noop {} {\bibfield  {journal} {\bibinfo  {journal}
  {Sov. {P}hys. {JETP}}\ }\textbf {\bibinfo {volume} {20}},\ \bibinfo {pages}
  {762} (\bibinfo {year} {1965})}\BibitemShut {NoStop}%
\end{thebibliography}%

\end{document}